\documentstyle[aps]{revtex}

\input{psfig}

\widetext

\begin{document}
\title{Quantum state protection in cavities}
\author{D. Vitali and P. Tombesi}
\address {Dipartimento di Matematica e Fisica, Universit\`a di
Camerino, via Madonna delle Carceri I-62032 Camerino \\
and Istituto Nazionale di Fisica della Materia, Camerino, Italy}
\author{G.J. Milburn}
\address{Physics Department, University of Queensland, St. Lucia, 4072, 
Brisbane, Australia}
\date{\today}
\maketitle

\begin{abstract}
We show how an initially prepared quantum state of a radiation mode in a 
cavity can be preserved for a long time using a feedback scheme 
based on the injection of appropriately prepared atoms. We present a 
feedback scheme both for optical cavities, which can be continuously 
monitored by a photodetector, and for microwave cavities, which can be 
monitored only indirectly via the detection of atoms that have interacted 
with the cavity field.
We also discuss the possibility of applying these methods for 
decoherence control in quantum information processing.
\end{abstract}


\section{Introduction}
Quantum optics is usually concerned with the generation of nonclassical 
states of the electromagnetic field and their experimental detection.
However with the recent rapid progress in the theory of quantum information
processing
the {\it protection} of quantum states and their quantum dynamics also is 
becoming a very important issue.
In fact what makes quantum information processing much more attractive 
than its classical counterpart is its capability of using entangled 
states and of processing generic linear 
superpositions of input states. The entanglement between a pair of 
systems is capable of connecting two observers separated by a space-like 
interval, it can neither be copied nor eavesdropped on without 
disturbance, nor can it be used by itself to send a classical message 
\cite{bennet}.
The possibility of using linear superposition states has given rise to 
quantum computation, which is essentially equivalent to have massive 
parallel computation \cite{eke}. 
However all these applications crucially 
rely on the possibility of maintaining quantum coherence, that is, a 
defined phase relationship between the different components of linear 
superposition states, over long distances and for long times. 
This means that one has to minimize as much as possible the effects of 
the interaction of the quantum system with its environment and, in 
particular, {\it decoherence}, i.e., the rapid destruction of the
phase relation between two quantum states of a system caused by the
entanglement of these two states with two different states of the
environment \cite{zur,leg1}.

Quantum 
optics is a natural candidate for the experimental implementation of 
quantum information processing systems, 
thanks to the recent achievements in the manipulation of 
single atoms, ions and single cavity modes. In fact two quantum gates 
have been already demonstrated \cite{turchette,wine1} in quantum 
optical systems and 
it would be very important to develop strategies capable of {\it 
controlling the decoherence} in experimental situations such 
as those described in Refs.~\cite{turchette,wine1}.

The possibility of an experimental control of decoherence is important 
also from a more fundamental point of view. In fact decoherence is the 
practical explanation of why linear superposition of
macroscopically distinguishable states, the states involved in the famous 
Schr\"odinger cat paradox \cite{cat}, are never observed and how the classical 
macroscopic world emerges from the quantum one \cite{zur}. 
In the case of macroscopic systems, the
interaction with the environment can never be escaped; since the
decoherence rate is proportional to the ``macroscopic separation''
between the two states \cite{zur,leg2,milwal}, a linear
superposition of macroscopically distinguishable states is
immediately changed into the corresponding statistical mixture, with
no quantum coherence left. Nonetheless, a full comprehension of the
fuzzy boundary between classical and quantum world is not yet
reached \cite{whee,tod}, and therefore the study of 
``Schr\"odinger cat'' states in {\it mesoscopic}
systems where one can hope to observe the decoherence is
important. A first achievement has been obtained by Monroe {\it et
al.} \cite{wine}, who have prepared a trapped ${\rm ^{9}Be^{+}}$ ion
in a superposition of spatially separated coherent states and
detected the quantum coherence between the two localized states.
However, in this experiment the decoherence of the 
superposition state has not been studied. 
The progressive decoherence of a mesoscopic Schr\"odinger cat has been 
observed for the first time in the experiment of Brune  
{\it et al}. \cite{prlha}, where the linear
superposition of two coherent states of the electromagnetic field
in a cavity with classically distinct phases has been generated and 
detected.

In this paper we propose a simple physical way to control decoherence
and protect a given quantum state against the destructive effects of the 
interaction with the environment: applying an appropriate feedback.
We shall consider a radiation mode in a cavity as the quantum system to 
protect and we shall show that the ``lifetime'' of an initial quantum 
state can be significantly increased and its quantum coherence properties 
preserved for quite a long time. The feedback scheme considered here 
has a quantum nature, since it is based on the injection of an 
appropriately prepared atom in the cavity and some preliminary aspects of the
scheme, and its 
performance, have been  described in Refs.~\cite{prlno,jmo}. The 
present paper is a much more detailed description of our approach to 
quantum state protection and is organized as follows. In Section II, the 
main idea is presented and a continuous feedback scheme for optical 
cavities is studied. In Section III, a possible application of this 
continuous feedback scheme to quantum 
information processing systems as the quantum phase gate of Ref.~
\cite{turchette} is presented. In the remaining sections, the stroboscopic 
version of the continuous feedback scheme, more suited for the microwave 
cavity of the Brune {\it et al.} experiment \cite{prlha}, and
first introduced in \cite{prlno}, is discussed in detail.  

\section{A feedback loop for optical cavities}

Applying a feedback loop to a quantum system means subjecting it to a 
series of measurements and then using the result of these 
measurements to modify the dynamics of the system. 
Very often the system is continuously monitored and 
the associated feedback scheme provides a continuous control of the 
quantum dynamics. An example is the measurement of an optical 
field mode, such as photodetection and homodyne measurements, and for 
these cases, Wiseman and Milburn have developed a quantum theory of 
continuous feedback \cite{feed}. This theory has been applied in 
Refs.~\cite{noi} to show that homodyne-mediated feedback
can be used to slow down the decoherence of a Schr\"odinger 
cat state in an optical cavity. 

Here we propose a different feedback scheme, 
based on direct photodection rather than homodyne 
detection. The idea is very simple: whenever the cavity looses a 
photon, a feedback loop supplies the cavity mode with another 
photon, through the injection of an appropriately prepared atom.
This kind of feedback is naturally suggested by the quantum 
trajectory picture of a decaying cavity field \cite{Carmichael}, in which 
time evolution is driven by the non-unitary evolution operator 
$\exp\{-\gamma t a^{\dagger}a/2\}$ interrupted at random times by an 
instantaneous jump describing the loss of a photon. 
The proposed feedback almost instantaneously ``cures'' the effect 
of a quantum jump and is able therefore to minimize the destructive effects of 
dissipation on the quantum state of the cavity mode.

In more general terms, the 
application of a feedback loop modifies the master 
equation of the system and therefore it is equivalent to an effective 
modification of the dissipative environment of the cavity field. For 
example, Ref.~\cite{squee} shows that a squeezed bath \cite{qnoise} 
can be simulated by the application of a feedback loop based on a 
quantum non-demolition (QND) measurement of a quadrature of a cavity 
mode. In other words, feedback is the main tool for realizing, 
in the optical domain, the so called ``quantum reservoir 
engineering'' \cite{poyatos}.  

The master equation for continuous feedback has been
derived by Wiseman and Milburn 
\cite{feed}, and, in the case of perfect detection via a single 
loss source, is given by
\begin{equation}
\dot{\rho }= \gamma \Phi (a \rho a^{\dagger}) -\frac{\gamma 
}{2}a^{\dagger}a \rho -\frac{\gamma }{2} \rho a^{\dagger}a \;,
\label{feedeq}
\end{equation}
where $\gamma $ is the cavity decay rate and $\Phi (\rho )$ is a generic 
superoperator describing the 
effect of the feedback atom  on the cavity state $\rho$.
Eq.~(\ref{feedeq}) assumes perfect detection, i.e., 
all the photons leaving the cavity are absorbed by a unit-efficiency
photodetector and trigger the cavity loop. It is practically impossible to
realize such an ideal situation and therefore it is more realistic to 
generalize this feedback master equation to the situation where
only a fraction  $\eta < 1$ of the photons leaking out of the cavity
is actually detected and switches on the atomic injector. It is 
immediate to see that (\ref{feedeq}) generalizes to
\begin{equation}
\dot{\rho }= \eta \gamma \Phi (a \rho a^{\dagger}) 
+(1-\eta) \gamma a \rho a^{\dagger}-\frac{\gamma 
}{2}a^{\dagger}a \rho -\frac{\gamma }{2} \rho a^{\dagger}a \;. 
\label{feedeq2}
\end{equation}

Now, we have to determine the action of the feedback atom on the cavity 
field $\Phi (\rho )$; this atom has to release exactly one photon 
in the cavity, possibly regardless of the field state in the cavity. In the 
optical domain this could be realized using
{\it adiabatic transfer of Zeeman coherence} \cite{adia}.

\subsection{Adiabatic passage in a three level $\Lambda $ atom}

A scheme based on the adiabatic passage of an atom with Zeeman 
substructure through overlapping cavity and laser fields has been 
proposed \cite{adia} for the generation of linear superpositions of Fock 
states in optical cavities. This technique allows for coherent 
superpositions of atomic ground state Zeeman sublevels to be ``mapped'' 
directly onto coherent superpositions of cavity-mode number states.
If one applies this scheme in the simplest case of a three-level $\Lambda 
$ atom one obtains just the feedback superoperator we are looking for, that is
\begin{equation}
\Phi (\rho ) = 
a^{\dagger} (aa^{\dagger})^{-1/2} \rho (aa^{\dagger})^{-1/2} a \;,
\label{adiafi} 
\end{equation}
corresponding to the feedback atom releasing exactly one photon into the cavity,
regardless the state of the field. 

\begin{figure}
\centerline{\psfig{figure=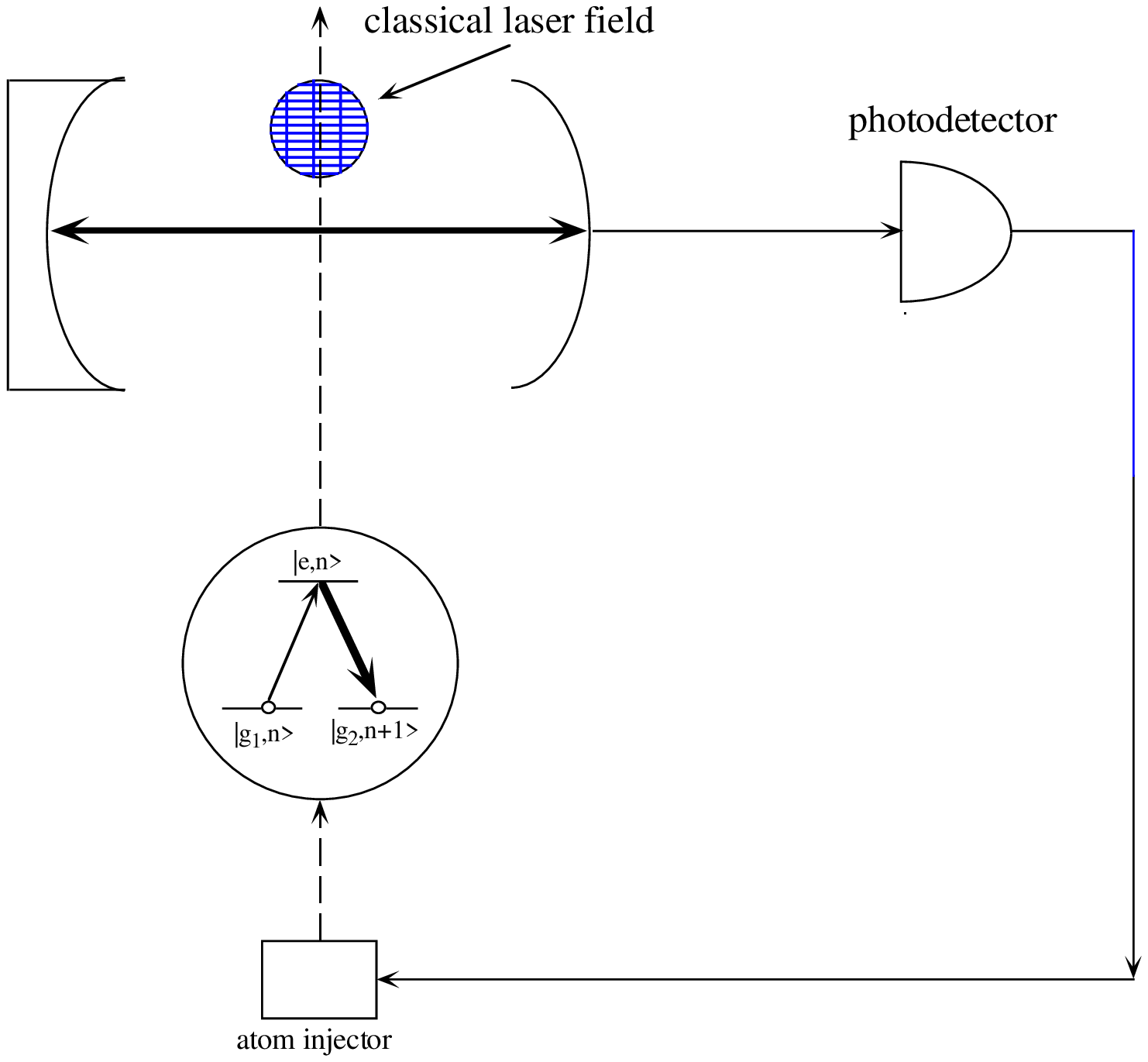,height=22cm}}
\caption{Schematic diagram of the photodetection-mediated feedback scheme
proposed for optical cavities, together with the appropriate atomic 
configuration for the adiabatic transfer.}
\label{appa1}
\end{figure}

To see this, let us consider
a three level $\Lambda$ atom with two ground states $|g_{1}\rangle$
and $|g_{2}\rangle$, coupled to the excited state $|e\rangle $ 
via, respectively, a classical laser field $\Omega (t)$ of frequency 
$\omega _{L}$, and a cavity field mode of frequency $\omega $. The 
corresponding Hamiltonian is
\begin{eqnarray}
&& H(t)= \hbar \omega a^{\dagger}a +\hbar \omega_{eg} |e\rangle \langle e|
-i\hbar g(t) \left(|e\rangle \langle g_{2}| a-|g_{2}\rangle \langle e| 
a^{\dagger}\right)  \nonumber  \\
&&+ i \hbar \Omega(t) \left(|e\rangle \langle 
g_{1}|e^{-i\omega _{L} t}-|g_{1}\rangle \langle e| e^{i \omega _{L} t}\right)
\;.
\label{jcm}
\end{eqnarray}
The time dependence of $\Omega(t)$ and $g(t)$ is provided by the 
motion of the atom across the laser and cavity profiles. This Hamiltonian 
couples only states within the three-dimensional manifold spanned by 
$|g_{1},n\rangle ,\,|e,n\rangle,\,|g_{2},n+1\rangle $, where $n$ denotes 
a Fock state of the cavity mode. Of particular interest within this 
manifold is the eigenstate corresponding to the adiabatic energy 
eigenvalue (in the frame rotating at the frequency $\omega $) 
$E_{n}=n\hbar \omega$,
\begin{equation}  
|E_{n}(t)\rangle = \frac{g(t) \sqrt{n+1}|g_{1},n\rangle  +\Omega(t) 
|g_{2},n+1 \rangle }{\sqrt{\Omega ^{2}(t)+(n+1) g^{2}(t)}}
\end{equation}
which does not contain any contribution from the excited state and for 
this reason is called the ``dark state''. This eigenstate exhibits the 
following asymptotic behavior as a function of time
\begin{equation}
|E_{n}\rangle \rightarrow \left\{\matrix{
                |g_{1},n\rangle & {\rm for} & \Omega(t)/g(t) \rightarrow 0  \cr
                |g_{2},n+1\rangle & {\rm for} & g(t)/\Omega(t) \rightarrow 0  \cr} \right.
\label{asymp}
\end{equation}
Now, according to the adiabatic theorem \cite{messiah}, 
when the evolution from time $t_{0}$ to time $t_{1}$ is sufficiently slow, a 
system starting from an eigenstate of $H(t_{0})$ will pass into the 
corresponding eigenstate of $H(t_{1})$ that derives from it by continuity.
This means that if the atom crossing is such that adiabaticity is satisfied,
when the atom enters the interaction region in the ground state 
$|g_{1}\rangle$, the following adiabatic transformation 
of the atom-cavity system state takes place
\begin{eqnarray}
\label{reali}
&& |g_{1}\rangle \langle g_{1}| \otimes \sum_{n,m}\rho _{n,m} 
|n\rangle \langle m| \\
&& \rightarrow |g_{2}\rangle \langle g_{2}| \otimes 
\sum_{n,m}\rho _{n,m} |n+1\rangle \langle m+1| \nonumber \\
&& = |g_{2}\rangle \langle g_{2}| \otimes 
a^{\dagger} (aa^{\dagger})^{-1/2} \rho (aa^{\dagger})^{-1/2} a \nonumber \;.
\end{eqnarray}
Roughly speaking, this transformation amounts to a single photon 
transfer from the classical laser field to the quantized cavity mode 
realized by the crossing atom, provided that a counterintuitive 
pulse sequence in which the classical laser field $\Omega (t)$ is 
time-delayed with respect to $g(t)$ is applied. Figure \ref{appa1} 
shows a simple diagram of the feedback scheme, together with the 
appropriate atomic configuration, cavity and laser field profiles 
needed for the adiabatic transformation considered. 

The quantitative conditions under which adiabaticity is satisfied are 
obtained from the requirement that the transition from the dark state 
$|E_{n}(t)\rangle $ to the other states be very small. One obtains 
\cite{adia,chim}
\begin{equation}
\Omega_{max},g_{max} \gg T_{cross}^{-1}  \;,
\end{equation}
where $T_{cross}$ is the cavity crossing time and $\Omega_{max},g_{max}$ 
are the two peak intensities. 

The above arguments completely neglect dissipative effects due to cavity 
losses and atomic spontaneous emission. For example, cavity dissipation 
couples a given manifold $|g_{1},n\rangle ,\,|e,n\rangle,\,|g_{2},n+1\rangle $
with those with a smaller number of photons. Since ideal adiabatic 
transfer occurs when the passage involves a single manifold, optimization 
is obtained when the photon leakage through the cavity is 
negligible during the atomic crossing, that is
\begin{equation}
T_{cross}^{-1} \gg  \bar{n} \gamma \;,
\end{equation}
where $\bar{n}$ is mean number of photons in the cavity. On the contrary, 
the technique of adiabatic passage is robust against the effects of 
spontaneous emission as, in principle, the excited atomic state 
$|e\rangle$ is never populated. Of course, in practice some fraction of 
the population does reach the excited state and hence large values of 
$g_{max}$ and $\Omega _{max}$ relative to the spontaneous emission rate 
$\gamma _{e}$ are desirable.
To summarize, the quantitative conditions for a practical realization of 
the adiabatic transformation (\ref{reali}) are
\begin{equation}
\Omega_{max},g_{max} \gg T_{cross}^{-1} \gg \bar{n} \gamma, 
\gamma_{e} \;,
\label{condi}
\end{equation}
which, as pointed out in \cite{adia}, could be realized in optical cavity QED
experiments. 

We note that when the adiabaticity conditions (\ref{condi}) are 
satisfied, then also the Markovian assumptions at the basis of the feedback 
master equation (\ref{feedeq2}) are automatically justified.
In fact, the continuous feedback theory of Ref.~\cite{feed} 
is a Markovian theory derived assuming that
the delay time associated to the 
feedback loop can be neglected with respect to the typical 
timescale of the cavity mode dynamics. In the present 
scheme the feedback delay time is due to the electronic trasmission time of the 
detection signal and, most importantly, by the interaction time $T_{cross} $ 
of the atoms with the field, while the typical timescale of the cavity field 
dynamics is $1/\gamma \bar{n}$. Therefore, the inequality on the right of
Eq.~(\ref{condi}) is essentially the condition for the validity of the Markovian
approximation and this {\it a posteriori} justifies our use of the Markovian 
feedback master equation (\ref{feedeq2}) from the beginning. 

\subsection{Properties of the adiabatic transfer feedback model}

When we insert the explicit expression (\ref{adiafi}) of the feedback 
superoperator into Eq.~(\ref{feedeq2}), the feedback master equation 
can be rewritten in the more transparent form 
\begin{equation}
\dot{\rho }= \frac{(1-\eta) \gamma}{2}\left(2 a \rho a^{\dagger}-
a^{\dagger}a \rho - \rho a^{\dagger}a \right)
-\frac{\eta \gamma}{2} \left [ \sqrt{\hat{n}},\left[\sqrt{\hat{n}},
\rho \right] \right]
\label{sqroot} 
\end{equation}
that is, a standard vacuum bath master equation with 
effective damping coefficient $(1-\eta) \gamma $ plus an unconventional 
phase diffusion term, in which the photon number operator is replaced by 
its square root and which can be called ``square root of phase diffusion''.

In the ideal case $\eta=1$, vacuum damping vanishes
and only the unconventional phase diffusion survives. As shown in 
Ref.~\cite{wise}, this is equivalent to say that ideal photodetection 
feedback is able to transform standard photodetection into a quantum 
non-demolition (QND) measurement of the photon number. In this ideal case,
a generic Fock state $|n\rangle$ is obviously preserved for an 
infinite time, since each photon lost by the cavity triggers the feedback 
loop which, in a negligible time, is able to give the photon back through 
adiabatic transfer. However, the photon injected by feedback 
has no phase relationship with the photons already present in the cavity 
and, as shown by (\ref{sqroot}), this results in phase diffusion. An 
alternative description of this phenomenon is that the photon 
injection process is essentially a nonlinear number amplifier which is 
necessarily accompanied by diffusion in the conjugate 
variable \cite{Wagner93}.
This means that feedback does not guarantee perfect state protection for 
a generic {\it superposition of number states}, even in 
the ideal condition $\eta =1$. In fact in this case, only the diagonal matrix 
elements in the Fock basis of the initial pure state are perfectly conserved, 
while the off-diagonal ones always decay to zero, ultimately leading 
to a phase-invariant state. However this does not mean that the 
proposed feedback scheme is good for preserving number states only, 
because the unconventional ``square-root of phase 
diffusion'' is much slower than the conventional one (described by a 
double commutator with the number operator). 

In fact the time evolution of a generic density matrix element in the 
case of feedback with ideal photodetection $\eta =1$ is
\begin{equation}
\rho _{n,m}(t)=\exp\left\{-\frac{\gamma 
t}{2}\left(\sqrt{n}-\sqrt{m}\right)^{2}\right\} \rho _{n,m}(0) \;,
\label{sqrt2}
\end{equation}
while the corresponding evolution in the presence of 
standard phase diffusion is
\begin{equation}
\rho _{n,m}(t)=\exp\left\{-\frac{\gamma 
t}{2}\left(n-m\right)^{2}\right\} \rho _{n,m}(0) \;.
\label{phadif}
\end{equation}

Since 
\begin{equation}
(n-m)^{2} \geq \left(\sqrt{n}-\sqrt{m}\right)^{2} = 
\frac{(n-m)^{2}}{\left(\sqrt{n}+\sqrt{m}\right)^{2}} \;\;\;\;\; 
\forall\;n,m \,
\label{ineq}
\end{equation}
each off-diagonal matrix element decays slower in the square root 
case and this means that the feedback-induced unconventional phase 
diffusion is slower than the conventional one. 

A semiclassical estimation 
of the diffusion constant can be obtained from the representation
of the master equation in terms of the Wigner function. When a generic 
state is expanded in the Fock basis as
\begin{equation}
\rho =\sum_{n,m}\rho _{n,m}|n\rangle \langle m| \;,
\end{equation}
the corresponding Wigner function is given by \cite{luc} (in polar 
coordinates $r,\theta $)
\begin{eqnarray}
\label{polar}
&& W(r,\theta )=\sum_{n}\rho _{n,n}\frac{2}{\pi }(-1)^n 
e^{-2r^2}L_{n}(4r^2) \\
&& + 2 {\rm Re}\left\{\sum_{n\neq m} \rho _{n,m}\frac{2}{\pi }(-1)^n 
\sqrt{\frac{n!}{m!}}e^{i\theta (m-n)}  \right. \nonumber \\
&& \times \left. (2r)^{m-n} e^{-2r^2}L^{m-n} _{n}(4r^2)
\right\} \;,\nonumber
\end{eqnarray}
where $L_{n}^{m-n}$ are the generalized Laguerre polynomials
and using this expression it is easy to see that
\begin{equation}
-\left[n,\left[n,\rho \right]\right] \leftrightarrow \frac{\partial 
^2}{\partial \theta ^2} W(r,\theta ) \;.
\label{derise}
\end{equation}
In the case of the square root of phase diffusion, one has instead
\begin{eqnarray}
\label{compli}
&&-\left[\sqrt{n},\left[\sqrt{n},\rho \right]\right] \leftrightarrow 
 2 {\rm
Re}\left\{\sum_{n\neq m} \rho _{n,m}\frac{2}{\pi }(-1)^n 
\sqrt{\frac{n!}{m!}} \right.  \\
&& \times \left.
\left(\sqrt{n}-\sqrt{m}\right)^{2}
e^{i\theta (m-n)} (2r)^{m-n} e^{-2r^2}L^{m-n} _{n}(4r^2)
\right\} \;;\nonumber
\end{eqnarray}
using (\ref{ineq}) and considering the semiclassical limit $n,m \gg 1$, 
$n \sim m \sim \bar{n}\gg 1$, where $\bar{n}$ is the mean photon number, 
(\ref{compli}) can be simplified to
\begin{equation}
-\left[\sqrt{n},\left[\sqrt{n},\rho \right]\right]
 \leftrightarrow \frac{1}{4 \bar{n}}\frac{\partial 
^2}{\partial \theta ^2} W(r,\theta ) \;,
\label{derise2}
\end{equation}
showing that (at least at large photon number)
in the case of the feedback-induced unconventional phase diffusion, the 
diffusion constant is scaled by a factor $4 \bar{n} \gg 1$.

A complementary description of the feedback-induced phase diffusion 
can be  
given by the time evolution of the mean coherent amplitude $\langle 
a(t) \rangle $. In fact, phase diffusion causes a decay of 
this amplitude as the phase spreads around  $2\pi $, even if the 
photon number is conserved.  In the presence of ordinary phase diffusion the
amplitude decays at the rate $\gamma /2$; in fact 
\begin{equation}
\langle a(t) \rangle = {\rm Tr}\left\{a \rho(t)\right\}=
\sum _{n=0}^{\infty}\sqrt{n+1} \rho_{n+1,n}(t) \;,
\label{amedio}
\end{equation}
and using Eq.~(\ref{phadif}) one gets
$$
\langle a(t) \rangle =e^{-\gamma t/2} \langle a(0) \rangle \;.
$$
In the case of the square root of phase diffusion,
Eqs.~(\ref{sqrt2}) and (\ref{amedio}) instead yield
\begin{equation}
\langle a(t) \rangle = {\rm Tr}\left\{a(t) \rho(0)\right\} \;,
\label{amedisq}
\end{equation}
where the Heisenberg-like time evolved amplitude operator $a(t)$ is 
given by
\begin{equation}
a(t) = \exp\left\{-\frac{\gamma t}{2}\left(\sqrt{a a 
^{\dagger}}-\sqrt{a^{\dagger} a}\right)^{2}\right\} a \;.
\label{amediti}
\end{equation}

In the semiclassical limit it is reasonable to assume
a complete factorization of averages 
(\ref{amedisq}), so to get
\begin{equation}
\langle a(t) \rangle = \exp\left\{-\frac{\gamma 
t}{2}\left(\sqrt{\bar{n}+1}-\sqrt{\bar{n}}\right)^{2}\right\} 
\langle a(0) \rangle \;,
\label{amediti2}
\end{equation}
which, in the limit of large mean photon number $\bar{n}$, yields a 
result analogous to that of Eq.~(\ref{derise2})
\begin{equation}
\langle a(t) \rangle = \exp\left\{-\frac{\gamma 
t}{8 \bar{n}}\right\} \langle a(0) \rangle \;.
\label{amediti3}
\end{equation}
This slowing down of phase diffusion (similar to that taking place 
in a laser well above threshold) means that, 
when the feedback efficiency $\eta $ is not too 
low, the ``lifetime'' of generic pure quantum states of
the cavity field can be significantly increased with respect to the 
standard case with no feedback (see Eq.~\ref{sqroot}).

\subsection{Description of the dynamics in the presence of feedback}

For a quantitative characterization of how the feedback scheme is able 
to protect an initial pure state we study the fidelity $F(t)$
\begin{equation}
F(t) = {\rm Tr}\left\{\rho(0) \rho(t)\right\}
\label{fido}
\end{equation}
i.e., the overlap between the final and the initial state $\rho(0)$ 
after a time $t$. In general $0\leq F(t)\leq 1$. For an initially pure state
$|\psi
(0)\rangle$, $F(t)$ is in fact the probability to find the system in the intial
state at a later
time. A decay to an asymptotic limit is given by the overlap $\langle \psi (0)
|\rho
(\infty)|\psi (0)\rangle$. 

A clear demonstration of the protection capabilities of the proposed 
feedback scheme is given when considering the preservation of initial
Schr\"odinger cat state,
i.e., the typical example of  nonclassical state whose oscillating and
non-positive definite Wigner 
function is a clear signature of quantum coherence \cite{zur}.
In fact, if the initial state is an even $(+)$ or odd $(-)$ 
Schr\"{o}dinger cat state
\begin{equation}
\label{eocat}
|\alpha_{\pm}\rangle= N_{\pm}\left 
(|\alpha\rangle \pm |-\alpha\rangle\right )
\end{equation}
where
\begin{equation}
N_{\pm}^{-2}=2\left(1\pm e^{-2 |\alpha |^{2}}\right) \;,
\label{npm}
\end{equation}
the corresponding fidelity $F(t)$ in absence of feedback ($\eta =0$ 
in (\ref{sqroot})) is given by
\begin{eqnarray}
&&F_{\pm}(t) = \frac{1+e^{-2 |\alpha |^{2}(1-e^{-\gamma t})}}{2}
e^{-|\alpha |^{2}(1-e^{-\gamma t/2})^{2}} \nonumber \\
&& \times \left(\frac{1\pm 
 e^{-2 |\alpha |^{2}e^{-\gamma t/2}}}{1\pm e^{-2 |\alpha 
|^{2}}}\right)^{2} \;.
\label{fideteo}
\end{eqnarray}

\begin{figure}
\centerline{\psfig{figure=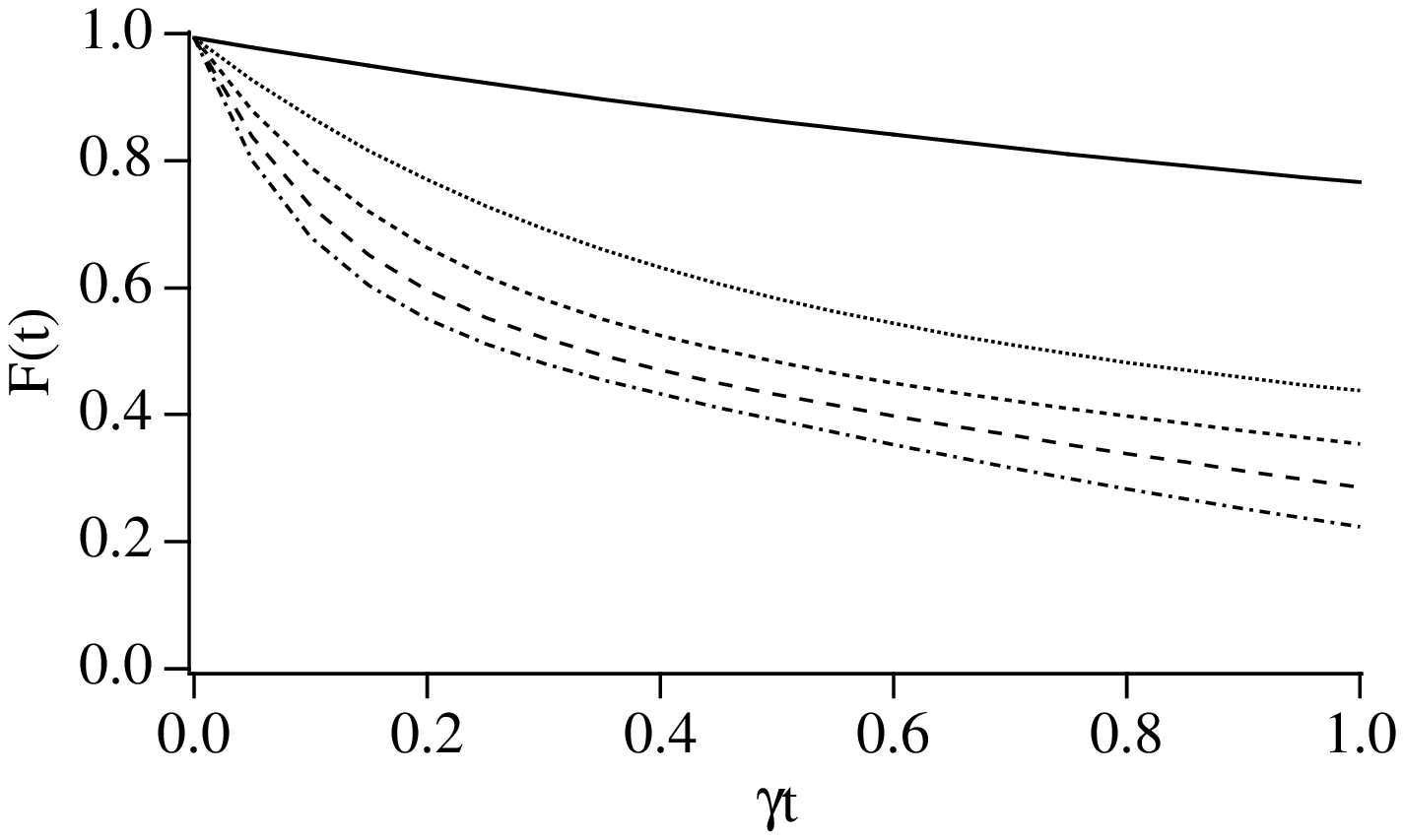,height=22cm}}
\caption{Time evolution of the fidelity $F(t)$ for an initial 
odd cat state with 
$|\alpha |^{2}= 5$; full line: $\eta =1$; dotted line: $\eta 
=0.75$; small dashes: $\eta =0.5$; big dashes: $\eta =0.25$;
dot-dashed line: evolution in absence 
of feedback ($\eta =0$).}
\label{fidecat}
\end{figure}

The corresponding function $F(t)$ in the presence of feedback can be 
easily obtained from the numerical solution of the master equation 
(\ref{sqroot}) and using the general expression
\begin{equation}
F(t) = \sum_{n,m}\rho_{n,m}^{*}(0) \rho_{n,m}(t) \;.
\label{fidege}
\end{equation}
The numerical results (Fig.~\ref{fidecat}) show that 
$F(t)$ in the presence of feedback is, at any time, significantly 
larger than the corresponding function in absence of feedback, even 
when the photodetection efficiency $\eta$ is far from the ideal 
value $\eta = 1$. 
Figure \ref{fidecat} refers to an initial odd cat state with 
$\alpha = \sqrt{5}$; the full line refers to the feedback model in the 
ideal case $\eta =1$; the dotted line to the feedback case with $\eta 
=0.75$, small dashes refer to the case $\eta =0.5$; big dashes 
refer to $\eta =0.25$ and the dot-dashed line to the evolution in absence 
of feedback ($\eta =0$). 
As 
expected, the preservation properties of the proposed scheme
worsen as the photedetection efficiency $\eta $ is decreased. 
Nonetheless, Figure \ref{fidecat} clearly shows how 
this photodetection-mediated feedback increases the ``lifetime'' of a 
generic pure state in the cavity, in the sense that the probability 
of finding the initial state at any time $t$ is larger 
than the corresponding probability in absence of feedback. 

\begin{figure}
\centerline{\psfig{figure=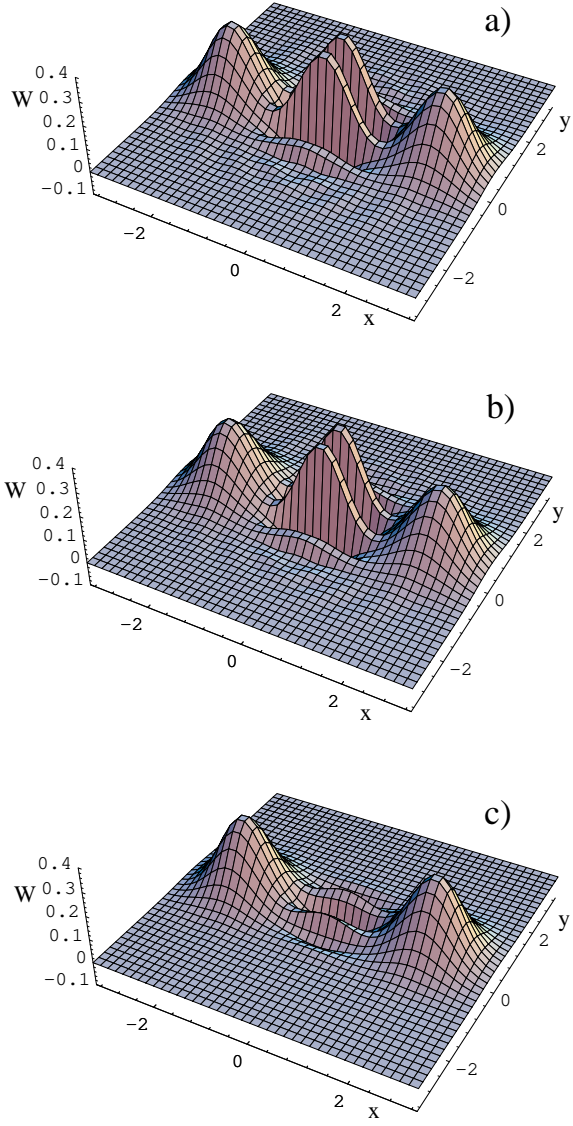,height=22cm}}
\caption{(a) Wigner function of the initial odd cat 
state, $|\psi \rangle= N_{-}(|\alpha \rangle - |-\alpha \rangle )$, $|\alpha 
|^{2}=5$; (b) Wigner function of the same cat state evolved for a 
time $t = 0.2/\gamma $ ($t=2 t_{dec}$),
in the presence of feedback ($\eta=1$); (c) 
Wigner function of the same state after a time $t=0.2/\gamma$, 
but evolved in absence of feedback.}
\label{wigcats}
\end{figure}

\begin{figure}
\centerline{\psfig{figure=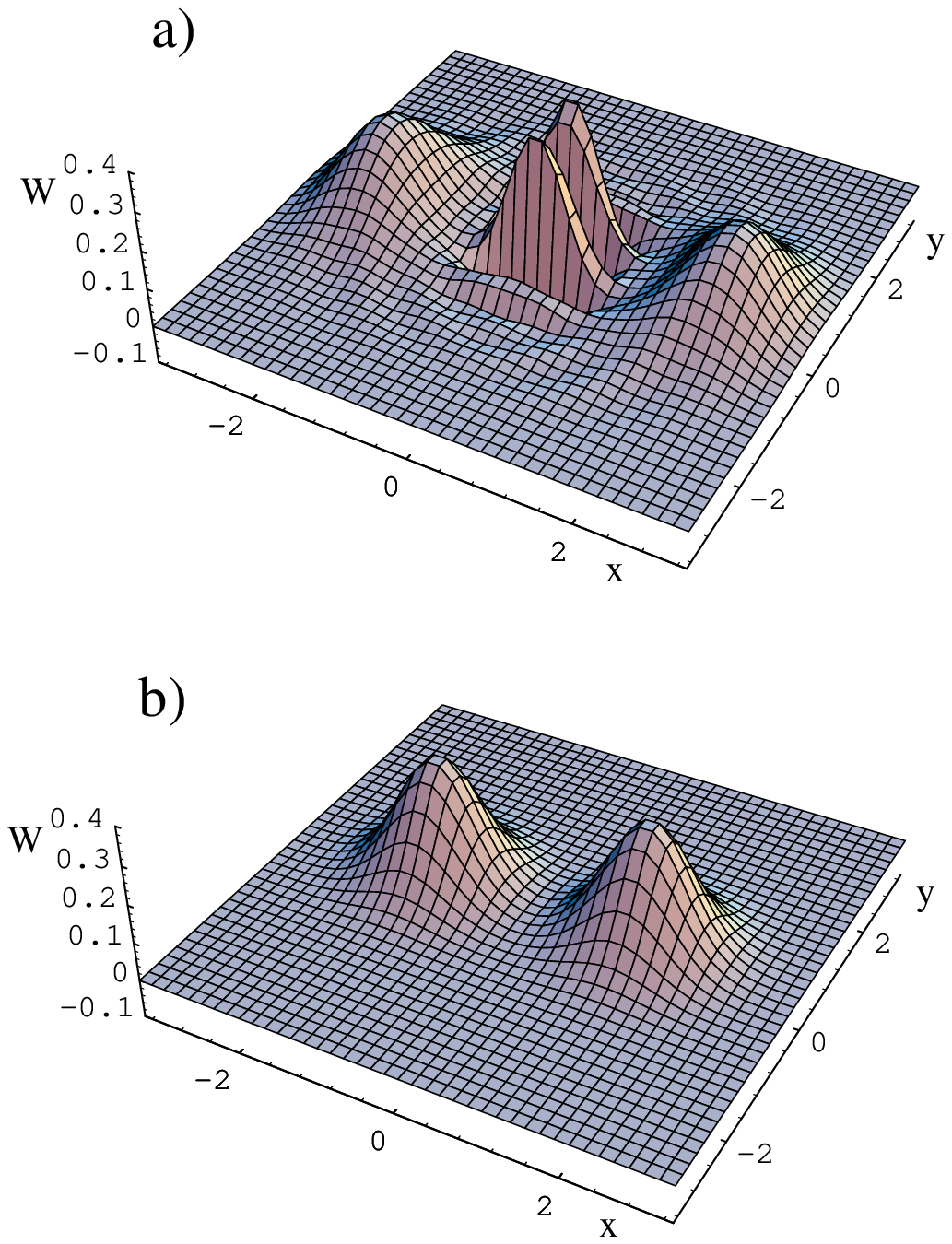,height=22cm}}
\caption{Wigner function of the odd cat state of Fig.~3,
evolved for a time $t = 1/\gamma $ in the presence of ideal
feedback $\eta=1$ (a), and in absence of feedback (b).}
\label{wigcats2}
\end{figure}

A qualitative confirmation of how well an initial odd cat state with 
$\alpha = \sqrt{5}$ is protected by feedback is given by Figure 
\ref{wigcats}: (a) shows the Wigner function of the initial cat 
state, (b) the Wigner function of the same cat state evolved for a 
time $t = 0.2/\gamma $ in the presence of feedback ($\eta=1$) and (c) the 
Wigner function of the same state again after a time $t=0.2/\gamma$, 
but evolved in absence of feedback. This elapsed time is twice the 
decoherence time of the Schr\"odinger cat state, $t_{dec}=(2 \gamma 
|\alpha |^{2})^{-1}$ \cite{leg2,milwal}, i.e., the lifetime of the 
interference terms in the cat state density matrix in the presence of the 
usual vacuum damping. As it is shown by (c), this means that after 
this short time the cat state has already lost the oscillating part 
of the Wigner function associated to quantum interference and has 
become a statistical mixture of two coherent states. This is no longer 
true in the presence of our feedback scheme: (b) shows that,
after $t \sim 2 t_{dec}$, the state is almost indistinguishable 
from the initial one and that the quantum wiggles of the Wigner function 
are still well visible. The capability of the feedback scheme of 
preserving the quantum coherence of the initial cat state for quite a 
long time is shown also by Fig.~\ref{wigcats2}, in which the Wigner 
function both in the presence ($\eta =1$) (a) and in absence of feedback 
(b) of the 
initial odd cat state of Fig.~\ref{wigcats} evolved after one relaxation 
time $t=\gamma ^{-1}$ is shown. In the presence of feedback the 
oscillating part between the two peaks is still visible, even if the 
state begins to be distorted with respect to the initial one because of 
the action of the unconventional phase diffusion which makes it more 
``rounded''. We obtain, so to speak, a ``mangy'' cat.

\begin{figure}
\centerline{\psfig{figure=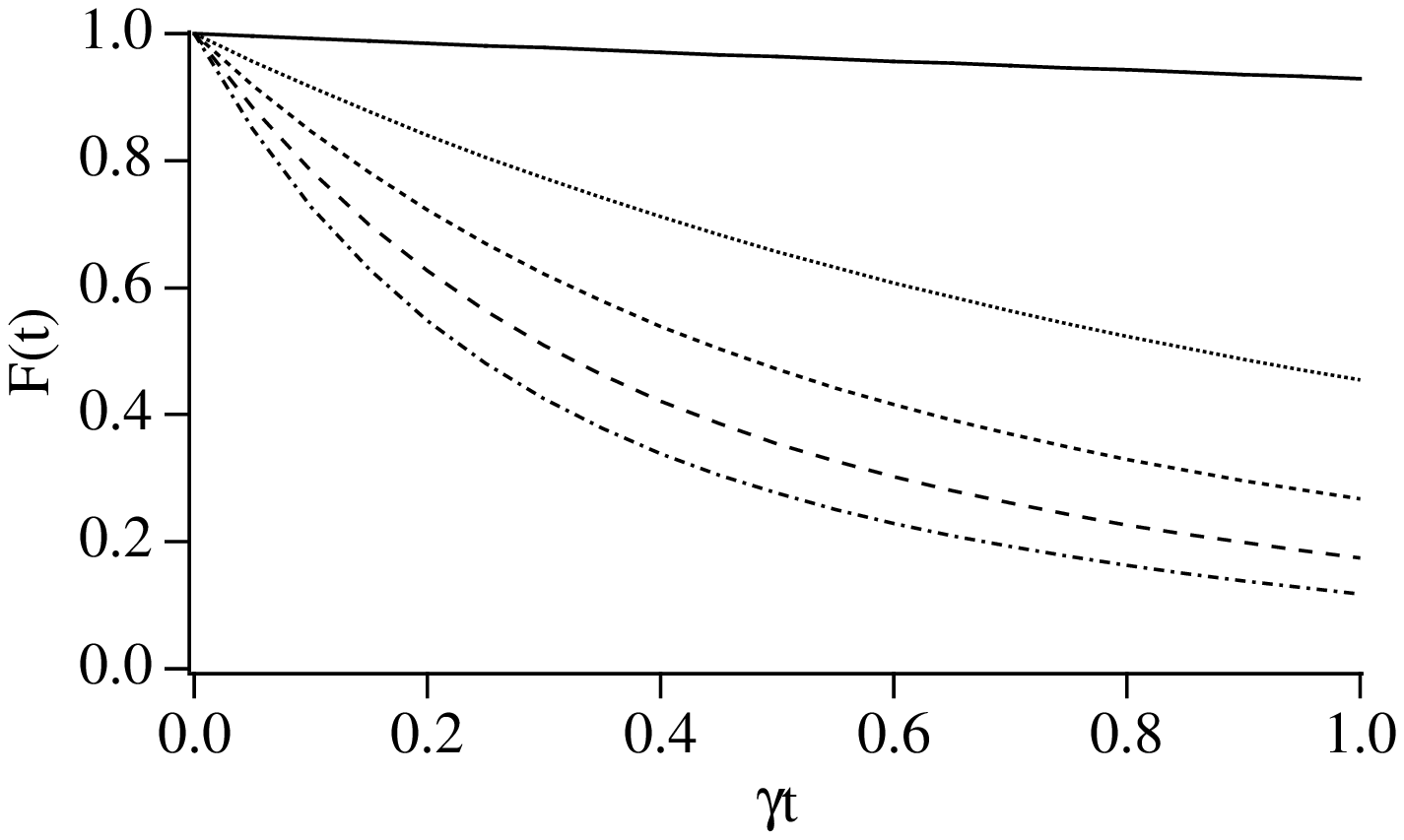,height=22cm}}
\caption{Time evolution of the fidelity $F(t)$ for the initial 
superposition state of two Fock states $(|2\rangle +2^{1/2} |4\rangle 
)/3^{1/2} $; full line: $\eta =1$; dotted line: $\eta 
=0.75$; small dashes: $\eta =0.5$; big dashes: $\eta =0.25$;
dot-dashed line: evolution in absence 
of feedback ($\eta =0$).}
\label{fidefok}
\end{figure}

Another clear example of how the quantum coherence associated to 
nonclassical superposition states of the radiation field inside the 
cavity is well preserved by the feedback scheme based on the adiabatic 
passage, is given by the study of the evolution of linear 
superpositions of two Fock number states 
\begin{equation}
|\psi(0)\rangle = \alpha |n\rangle +\beta |m \rangle \;.
\label{foklin}
\end{equation}
These states have not been experimentally generated in optical 
cavities yet, but there are now a number of proposals for their generation
\cite{fok,eberly}. In this case $F(t)$ can be easily evaluated analytically
($m>n$)
\begin{eqnarray}
&& F(t) = |\alpha|^{4}e^{-n (1-\eta)\gamma t}+|\beta|^{4}e^{-m (1-\eta)\gamma t}
\nonumber \\
&&+2 |\alpha|^{2} |\beta|^{2} e^{-\gamma t \left[(m+n)/2-\eta \sqrt{n 
m}\right]}  \\
&& +|\alpha|^{2} |\beta|^{2} e^{-n (1-\eta)\gamma t} 
\left(1-e^{-(1-\eta)\gamma t}\right)^{m-n}\frac{m! n!}{(m-n)!} \nonumber
\end{eqnarray}
and when this expression is plotted for different values of $\eta$ 
and compared with that in absence of feedback ($\eta =0$), we see, as 
in Figure \ref{fidecat}, a significant increase of the ``lifetime'' of 
the state (\ref{foklin}).
This comparison is shown in Figure \ref{fidefok},
which refers to the initial state $(|2 \rangle +\sqrt{2} |4\rangle 
)/\sqrt{3}$  and where the notation is as in Fig.~\ref{fidecat}:
the full line refers to the feedback model in the 
ideal case $\eta =1$; the dotted line to the feedback case with $\eta 
=0.75$, small dashes refer to the case $\eta =0.5$; big dashes 
refer to $\eta =0.25$ and the dot-dashed line to the evolution in absence 
of feedback ($\eta =0$). 

\section{Optical feedback scheme for the protection of qubits}

Photon states are known to retain their phase coherence over considerable 
distances and for long times and for this reason high-Q optical cavities 
have been proposed as a promising example for the realization of simple 
quantum circuits for quantum information processing. To act as an 
information carrying quantum state, the electromagnetic fields must 
consist of a superposition of few distinguishable states. The most 
straightforward choice is to consider the superposition of the vacuum and 
the one photon state $\alpha |0\rangle +\beta |1\rangle $. However it is easy 
to understand that this is not convenient because any interaction 
coupling $|0\rangle $ and $|1\rangle $ also couples $|1\rangle $ with 
states with more photons and this leads to information losses. Moreover 
the vacuum state is not easy to observe because it cannot be 
distinguished from a failed detection of the one photon state. A more 
convenient and natural choice is {\it polarization coding}, i.e., using 
two degenerate polarized modes and qubits of the following form
\begin{equation}
|\psi \rangle = \left(\alpha a_{+}^{\dagger}+\beta a_{-}^{\dagger}\right) 
|0\rangle = \alpha |0,1\rangle + \beta |1,0\rangle \;,
\label{twoqub}
\end{equation}
in which one photon is shared by the two modes \cite{sten}.
In fact this is a ``natural'' two-state system, in which the two basis 
states can be easily distinguished with polarization measurements; 
moreover they can be easily transformed into each other using polarizers.
Polarization coding has been already employed in one of the few 
experimental realization of a quantum gate, the quantum phase gate 
realized at Caltech \cite{turchette}. This experiment has demonstrated 
conditional quantum dynamics between two frequency-distinct fields
in a high-finesse optical cavity. The implementation of this gate 
employs two single-photon pulses with frequency separation large 
compared to the individual bandwidth, and whose internal state is 
specified by the circular polarization basis as in (\ref{twoqub}).
The conditional dynamics between the two fields is obtained 
through an effective strong 
Kerr-type nonlinearity provided by a beam of cesium atoms. 

In the preceding section we have shown that the proposed feedback scheme 
is able to increase the ``lifetime'' of linear 
superpositions of Fock states. Therefore it is quite natural to look 
if our scheme can be used to protect qubits like those of the Caltech 
experiment, against the destructive effects of cavity damping. To be more 
specific, here we shall not be concerned with the protection of the 
quantum gate dynamics, but we shall focus on a simpler but still 
important problem: protecting an unknown input state for the longest 
possible time against decoherence. For this reason we shall not consider 
the two interacting fields, but a single frequency mode with a generic 
polarization, i.e., a single qubit. We shall consider a class of 
initial states more general than those of Eq.~(\ref{twoqub}), i.e., 
\begin{equation}
|\psi \rangle =  \alpha |n,m\rangle + \beta |m,n\rangle \;,
\label{twoqubgen}
\end{equation}
where $m+n$ photons are shared by the two polarized modes.

If we want to apply the adiabatic transfer feedback scheme described 
above for protecting qubits as those of 
Eq.~(\ref{twoqubgen}), one has to consider a feedback loop 
as that of Fig.~1 for each polarized mode. This can be done 
using polarization-sensitive detectors which electronically control
the polarization of the classical laser field and the initial state 
of the injected atoms. In fact one has to release in the cavity
a left or right 
circularly polarized photon depending on which detector has fired
and this can be easily achieved when the $|g_{1}\rangle \rightarrow |e\rangle $
and $|g_{2}\rangle \rightarrow |e\rangle $ transitions are characterized
by opposite angular momentum difference
$\Delta m_{J} = \pm 1$. In this case a left polarized photon,
for example, is given back to the cavity with the adiabatic
transition $|g_{1}\rangle \rightarrow |g_{2}\rangle $ of Fig.~1, while 
the right polarized one is released into the cavity
through the reversed adiabatic transition 
$|g_{2}\rangle \rightarrow |g_{1}\rangle $ and the two possibilities 
are controlled by the polarization sensitive detectors.

Since the input state we seek to protect is unknown, the protection capabilities
of the feedback scheme are better characterized by the minimum fidelity,
i.e., the fidelity of Eq.~(\ref{fido}) minimized over all 
possible initial states. This minimum fidelity can be easily evaluated 
by solving the master equation (\ref{sqroot}) for each polarized
mode and one gets the following expression
\begin{equation}
F_{min}(t)=\frac{1}{2}\left(
e^{-(1-\eta )\gamma t (n+m)}+e^{-\gamma t (n+m-2\eta \sqrt{n m})}\right) 
\;.
\label{minfid}
\end{equation}

In the absence of feedback ($\eta =0$), this expression becomes 
$F_{min}(t)=\exp\{-\gamma t(n+m)\}$ showing that in this case, the 
states  most robust against cavity damping are those with the smallest number 
of photons, $m+n=1$, i.e., the states of the form of Eq.~(\ref{twoqub}).
Moreover, in a typical quantum information processing situation, one has 
to consider small qubit ``storage'' times $t$ with respect to $\gamma 
^{-1}$ so to have reasonably small error probabilities in quantum 
information storage. Therefore the protection capability of an optical 
cavity with no feedback applied is described by 
\begin{equation}
F_{min}(t) = 1-\gamma t \;.
\label{nofed}
\end{equation}

If we now consider the situation in the presence of feedback 
(Eq.~(\ref{minfid})), the best protected states for a given nonzero 
efficiency $\eta $, may be different from the states with only one 
photon, $\alpha |0,1\rangle+\beta 
|1,0\rangle$, and they depend upon the explicit value of the feedback efficiency 
$\eta$. For the determination of the optimal qubit of the form of 
(\ref{twoqubgen}) (i.e., the optimal values for $m$ and $n$), one has to 
minimize the deviation from the perfect protection condition $F(t)=1$.
For $\gamma t \ll 1$ one gets 
\begin{eqnarray}
&&\min_{m \neq n} \left[(2-\eta)(n+m)-2\eta \sqrt{nm}\right] 
\nonumber \\
&& =\min_{m \neq n} \left[(\sqrt{m}-\sqrt{n})^{2}+(1-\eta )
(\sqrt{m}+\sqrt{n})^{2}\right] \\
&& =\min_{n \geq 0,\, p \geq 1} \left[\frac{p^{2}}{(\sqrt{n+p}+\sqrt{n})^{2}}
+(1-\eta )
(\sqrt{n+p}+\sqrt{n})^{2}\right] \nonumber \;,
\end{eqnarray} 
where $p=m-n$.
From these expression it can be easily seen that one has 
to choose $p=1$, and 
therefore the optimal qubits are those of the form
\begin{equation}
|\psi \rangle =\alpha |n_{opt},n_{opt}+1\rangle +\beta 
|n_{opt}+1,n_{opt}\rangle \;,
\label{larpho}
\end{equation}
where $n_{opt}$ is determined by the minimization condition
\begin{equation}
\min_{n \geq 0} \left[\frac{1}{(\sqrt{n+1}+\sqrt{n})^{2}}
+(1-\eta )
(\sqrt{n+1}+\sqrt{n})^{2}\right] \;.
\label{mincond}
\end{equation} 
As long as 
\begin{equation}
\eta \leq 2/(1+\sqrt{2}) \simeq 0.83 \;,
\label{p83}
\end{equation}
one has 
$n_{opt}=0$ and therefore the situation is similar to that of the 
no-feedback case: the states of the form (\ref{twoqub}) are the best 
protected states
and the corresponding minimum fidelity is given by
\begin{equation}
F_{min}(t)=\frac{1}{2}\left(
e^{-(1-\eta )\gamma t}+e^{-\gamma t }\right) \simeq 1-\gamma t 
\left(1-\frac{\eta }{2}\right) \;.
\label{minfid2}
\end{equation}
In this case, feedback leads to a very poor qubit protection
with respect to the no-feedback case and therefore our scheme proves 
to be practically useless for the protection of single photon qubits of 
(\ref{twoqub}) employed in the Caltech experiment of 
Ref.~\cite{turchette}.

However, when the feedback efficiency $\eta$ becomes larger than 0.83,
the situation can improve considerably. In fact $n_{opt}$ becomes 
nonzero and can become very large in the limit $\eta \rightarrow 1$, 
and in this case the minimum fidelity decays very slowly. To be more 
specific, $n_{opt}$ is approximately given by the condition
\begin{equation}
\left(\sqrt{n_{opt}+1}+\sqrt{n_{opt}}\right)^{2}=\left(1-\eta\right)^{-1/2}
\label{fgfg}
\end{equation}
and the corresponding small time behavior of $F_{min}(t)$ is given by 
\begin{equation}
F_{min}(t) \simeq   1-\frac{\gamma t}{\left(\sqrt{n_{opt}+1}+\sqrt{n_{opt}}
\right)^{2}} \simeq 1- \gamma t \sqrt{1-\eta} \;.
\end{equation}
This means that in the limit of a feedback efficiency very close to 
one, it becomes convenient to work with a large number of photons per 
mode, since in this limit the probability of errors in the storage of quantum 
information can be made very small. This can be easily understood 
from Eq.~(\ref{sqroot}), because in this limit the square-root of 
phase diffusion term prevails in the master equation and its quantum 
state protection capabilities improve for increasing photon number 
(see Eq.~(\ref{amediti3})).
In the ideal case $\eta =1$, $n_{opt}$ becomes infinite and therefore  
the minimum fidelity can remain arbitrarily close to one.
It is convenient to work with the largest possible number of photons, 
that is,
\begin{equation}
|\psi \rangle =\alpha |n,n+1\rangle +\beta |n+1,n\rangle \;\;\;\;\; n \gg1
\label{larpho2}
\end{equation}
and the corresponding minimum fidelity is
$$
F_{min}(t)\simeq  \frac{1}{2}\left(1+ 
e^{-\gamma t/4n}\right) \simeq 1-\frac{\gamma t}{8n} \;.
$$

The feedback method proposed here to deal with decoherence in quantum 
information processing is different from most of the proposals made in 
this research field, which are based
on the so called quantum error correction codes \cite{error}, which 
are a way to use {\it software} to preserve linear superposition states. 
In our case, feedback allows a physical control of decoherence, through a 
continuous monitoring and eventual correction of the dynamics and in this 
sense our approach is similar in spirit to the approach of 
Ref.~\cite{pell,mabuchi}.
The present feedback scheme
is not very useful in the case of one-photon qubits
(\ref{twoqub}) of the quantum phase gate experiment of 
Ref.~\cite{turchette}; however it predicts a very good decoherence 
control in the case of high feedback efficiency $\eta > 0.83$ and for 
larger photon numbers (see Eq.~(\ref{larpho})). It is very difficult 
to achieve these experimental conditions with the present technology, 
but our scheme could become very promising in the future.

\section{A feedback scheme for microwave cavities}

In the case of measurements of an optical 
field mode, such as photodetection and homodyne measurements, the system 
is {\it continuously} measured and in these cases applying a feedback loop 
can be quite effective in controlling the decoherence of an 
optical Schr\"odinger cat. 
It is therefore quite natural to see if a similar control of 
decoherence can be achieved in the only (up to now) experimental
generation and detection of Schr\"odinger cat states of a radiation mode,
the experiment of Brune {\it et al} \cite{prlha}. However, in this experiment,
it is not possible
to  monitor continuously the state of the radiation in the cavity, since 
the involved field is in the microwave range and there are not good 
enough detectors in this wavelength region. 
The detection of the cat state is obtained through 
measurements performed on a second probe atom crossing the cavity after a 
delay time $T$ and that provides a sort of impulsive measurements of the 
cavity field state.

This suggests that in this microwave case,
continuous measurement can be replaced 
at best by a series of {\it repeated} measurements, performed by 
off-resonance atoms crossing the high-Q microwave cavity one by one with a 
time interval $T$. As a consequence, one could try to apply a sort of 
``discrete'' feedback scheme modifying in a ``stroboscopic''
way the cavity field dynamics according to the result of the 
atomic detection. 

\subsection{A simplified description of the experiment of 
Brune et al.}

In Ref.~\cite{prlha}, a Schr\"odinger cat state for the microwave field 
in a superconducting cavity $C$ has been generated using circular Rydberg 
atoms crossing the cavity in which a coherent state has been 
previously injected. 
All the atoms have an appropriately selected velocity and the relevant 
levels are two adjacent circular Rydberg states with principal 
quantum numbers $n=50$ and $n=51$, which 
we denote as $|g\rangle$ and $|e\rangle$ respectively. These two 
states have a very long lifetime ($30$ ms) and a very strong coupling 
to the radiation and 
the atoms are initially prepared in the state $|e\rangle$. 
The high-Q superconducting cavity is sandwiched between two low-Q 
cavities $R_{1}$ and $R_{2}$, 
in which classical microwave fields can be applied and which are 
resonant with the transition between the state $|e\rangle$ and the 
nearby lower circular state $|g\rangle$. The 
intensity of the field in the first cavity $R_{1}$ 
is then chosen so that, for the 
selected atom velocity, a $\pi/2$ pulse is applied to the atom as it 
crosses $R_{1}$. As a consequence, the atomic state before entering 
the cavity $C$ is 
\begin{equation}
|\psi_{atom}\rangle = \frac{1}{\sqrt{2}}\left(|e\rangle 
+|g\rangle\right) \;.
\label{at1}
\end{equation}

The high-Q cavity $C$ is slightly off-resonance with respect to the $e\, 
\rightarrow \, g$ transition, with detuning
\begin{equation}
\delta = \omega - \omega_{eg}\;,
\label{detu}
\end{equation}
where $\omega $ is the cavity mode frequency and 
$\omega_{eg}=(E_{e}-E_{g})/\hbar $. The Hamiltonian of the 
atom-microwave cavity mode system is the usual Jaynes-Cummings Hamiltonian,
given by
\begin{eqnarray}
&&H_{JC}=E_{e}|e\rangle \langle e | + E_{g}|g\rangle \langle g |+
\hbar \omega a^{\dagger} a \nonumber \\
&&+\hbar \Omega \left(|e \rangle \langle g |a+|g \rangle \langle e |
a^{\dagger}\right) \;,
\label{jc}
\end{eqnarray}
where $\Omega $ is the vacuum Rabi coupling between the atomic dipole 
on the $e\, \rightarrow \, g$ transition and the cavity mode.
In the off-resonant case and perturbative limit $\Omega \ll \delta $, the 
atom and the field essentially do not exchange energy 
but only undergo dispersive 
frequency shifts depending on the atomic level \cite{harray,brune}, and 
the Hamiltonian (\ref{jc}) becomes equivalent to
\begin{eqnarray}
&&H_{disp}=E_{e}|e\rangle \langle e | + E_{g}|g\rangle \langle g |+
\hbar \omega a^{\dagger} a \nonumber \\
&&+\hbar \frac{\Omega ^{2}}{\delta} \left(|g \rangle \langle g 
|a^{\dagger}a
-|e \rangle \langle e | a a^{\dagger}\right) \nonumber \\
&&=\left(E_{e}-\hbar \frac{\Omega ^{2}}{\delta}\right)
|e\rangle \langle e | + E_{g}|g\rangle \langle g |+
\hbar \left(\omega +\frac{\Omega ^{2}}{\delta}\right)
a^{\dagger} a \nonumber \\
&&-2\hbar \frac{\Omega ^{2}}{\delta} |e \rangle \langle e | 
a^{\dagger} a \;.
\label{heff}
\end{eqnarray}
This means that in this dispersive limit, besides a negligible shift 
of the cavity frequency and of the $e$ level energy, the atom-field
interaction induces a phase shift $\phi = 2 \Omega ^{2}t_{int}/\delta $
when the atom is in the state $e$, while there is no shift when the 
atom is in the state $g$ ($t_{int}$ is the interaction time).
Therefore, using (\ref{at1}), the state of the atom-field system when 
the atom has just exited the cavity $C$ is the entangled state
\begin{equation}
|\psi_{atom+field}\rangle = \frac{1}{\sqrt{2}}\left(|e,\alpha e^{i\phi}
\rangle 
+|g,\alpha \rangle\right) \;,
\label{atf1}
\end{equation}
where $\alpha$ denotes the coherent state initially present within 
the cavity. In the experiment of Ref.~\cite{prlha}, different values 
of the phase shift $\phi $ 
have been considered; however we shall restrict from now on
to the case $\phi=\pi$, which corresponds to the generation of a linear 
superposition of two coherent states with {\it opposite} phases.

In the state (\ref{atf1}), each atomic state is correlated to a different 
field phase; for the generation of a cat state, however, one has to 
correlate each atomic state to a {\it superposition} of coherent 
states with different phases, and this is achieved by submitting the 
atom to a second $\pi/2$ pulse in the second microwave cavity $R_{2}$. 
The $\pi/2$ pulse yields the following transformation
\begin{eqnarray}
|e\rangle & \rightarrow & \frac{1}{\sqrt{2}}\left(|e\rangle 
+|g\rangle\right) \nonumber \\
|g\rangle & \rightarrow & \frac{1}{\sqrt{2}}\left(-|e\rangle 
+|g\rangle\right) \;,
\label{pi2}
\end{eqnarray}
so that the state (\ref{atf1}) becomes
\begin{equation}
|\psi'_{atom+field}\rangle = 
\frac{1}{\sqrt{2}}\left(N_{-}^{-1}|e\rangle|\alpha_-\rangle 
+ N_{+}^{-1}|g\rangle |\alpha_+\rangle\right) \;.
\label{atf2}
\end{equation}
where $|\alpha _{\pm}\rangle $ are the even $(+)$ or odd $(-)$ Schr\"{o}dinger
cat states defined in (\ref{eocat}) and $N_{\pm}$ are defined in 
Eq.~(\ref{npm}).
Eq.~(\ref{atf2}) shows that
an even or an odd coherent state is conditionally generated in the 
cavity according to whether or not the atom is detected in the level $|g\rangle$
or $|e\rangle$, respectively.

After generation, the Schr\"odinger cat state undergoes 
a vary fast decoherence process \cite{leg2,milwal}, 
that is, a fast decay of
interference terms, caused by the inevitable presence of 
dissipation in the superconducting cavity. In fact the dissipative 
time evolution of the generated cat state is described by the 
following density matrix
\begin{eqnarray}
&&\rho(t)=\frac{1}{N_{\pm}^{2}}\left[|\alpha e^{-\gamma t/2}\rangle 
\langle \alpha e^{-\gamma t/2}|+|-\alpha e^{-\gamma t/2}\rangle 
\langle -\alpha e^{-\gamma t/2}| \right. \nonumber \\
&&\left.\pm e^{-2|\alpha|^{2}(1-e^{-\gamma 
t})}\left(|-\alpha e^{-\gamma t/2}\rangle 
\langle \alpha e^{-\gamma t/2}| \right. \right. \nonumber \\
&& \left. \left. +|\alpha e^{-\gamma t/2}\rangle 
\langle -\alpha e^{-\gamma t/2}|\right)\right] \;,
\label{dec}
\end{eqnarray}
where
$\gamma$ is the cavity decay rate and where the plus (minus) sign
correspond to the even (odd) coherent state. Decoherence is governed by 
the factor $\exp\left[-2|\alpha|^{2}(1-e^{-\gamma 
t})\right]$, which for $\gamma t \ll 1$ becomes 
$\exp\left[-2|\alpha|^{2}\gamma t\right]$, implying therefore that the 
interference terms decay to zero with a 
lifetime $t_{dec}=(2\gamma |\alpha |^{2})^{-1}$. 

The relevance of the experiment of Brune {\it et al}. \cite{prlha} 
lies in the fact that this progressive decoherence of the cat state
has been observed for the first time and the theoretical prediction 
checked with no fitting parameters.  
This monitoring of decoherence has been obtained by sending a 
second atom through 
the same arrangements of cavities. The atom has exactly the same 
velocity of the first atom generating the cat and is sent through the 
cavities after a time delay $T$, which is much larger 
than the time of flight of the atom through the whole system (which is 
of the order of $10^{-5} s$ in the experiment).
The state of the system composed by the second atom and the microwave 
field undergoes the same transformation described above for the first 
Rydberg atom, i.e., 
\begin{eqnarray}
\label{2at2}
&&\rho_{atom+field} \\
&&=U_{\frac{\pi}{2}}e^{i
\pi a^{\dagger}a |e\rangle \langle e|}U_{\frac{\pi}{2}}(\rho(T) \otimes
|e\rangle \langle e|) U_{\frac{\pi}{2}}^{\dagger} 
e^{-i\pi a^{\dagger}a |e\rangle \langle 
e|}U_{\frac{\pi}{2}}^{\dagger} \nonumber \;,
\end{eqnarray}
where $U_{\frac{\pi}{2}}$ describes the $\pi/2$ pulse and
$\rho(T)$ is the cavity field at a time $T$ after the passage 
of the first atom and it is given by Eq.~(\ref{dec}).

Using (\ref{pi2}) one finally gets the state of the probe atom+field 
system just before the field ionization detectors for the measurement 
of the $e$ or $g$ atomic state, that is, 
\begin{eqnarray}
&&\rho_{atom+field}= |e\rangle \langle e| \otimes \rho_{e} +
|g\rangle \langle g| \otimes \rho_{g} \nonumber \\
&& +|e\rangle \langle g| \otimes \rho_{+} +
|g\rangle \langle e| \otimes \rho_{-} \;,
\label{2at3}
\end{eqnarray}
where 
\begin{eqnarray}
\rho_{\frac{g}{e}}&=&\frac{1}{4}\left[P\rho P + \rho \pm P \rho \pm \rho 
P\right]
\label{offre} \\
\rho_{\pm}&=&\frac{1}{4}\left[P\rho P - \rho \pm P \rho \mp \rho 
P\right] \;,
\end{eqnarray}
and
\begin{equation}
P = e^{\pm i \pi a ^{\dagger}a}
\label{pari}
\end{equation}
is the parity operator of the microwave cavity mode .
From these expressions, the probability of detecting the second atom in 
the $e$ or $g$ state is readily obtained
\begin{equation}
P_{\frac{g}{e}}= \frac{1}{2}\left(1\pm \langle P \rangle \right) \;,
\label{prob}
\end{equation}
where $\langle P \rangle $ is the mean value of the
parity of the cavity mode state $\rho(T)$.
If one inserts in (\ref{prob}) the explicit expression of $\rho(T)$ 
given by (\ref{dec}), one gets the four conditional probabilities 
$P_{ij}$, ($i,j=e$ or $g$), of detecting the second atom in the state 
$j$ after detecting the first atom in the state $i$ and which 
give a satisfactory description of the decoherence process of 
the cat state in the cavity \cite{dav}. 
Let us consider for example the case of 
two successive detections of the circular Rydberg state $e$: in this 
case the detection of the first atom projects the microwave field in 
the superconducting cavity in an odd coherent state and the 
corresponding conditional probability is given by
\begin{equation}
P_{ee}(T)=\frac{1}{2}\left[1-\frac{e^{-2|\alpha |^{2}e^{-\gamma T}}-
e^{-2|\alpha |^{2}\left(1-e^{-\gamma T}\right)}}{1-e^{-2|\alpha 
|^{2}}}\right] \;.
\label{pee}
\end{equation}

The dependence of this conditional probability upon the time delay 
between the two atom crossings gives a clear description of the cat 
state decoherence. In fact, if there is no dissipation in the cavity,
i.e., $\gamma T =0$, it is $P_{ee}=1$ and this perfect correlation 
between the atomic state and the cavity state is the experimental 
signature of the presence of an odd coherent state in the high-Q 
cavity. As long as $\gamma \neq 0$, the conditional probability 
decreases for increasing delay time $T$. At a first stage one has a 
decay to the value $P_{ee}=1/2$ in the decoherence time 
$t_{dec}=1/2\gamma |\alpha|^{2}$; this is the decoherence process 
itself, that is, the fast transition from the quantum 
linear superposition state to the statistical mixture 
\begin{equation}
\rho_{mixt}=\frac{1}{2}\left[|\alpha \rangle \langle \alpha |+
|-\alpha \rangle \langle -\alpha |\right]
\label{mixt}
\end{equation}
describing a {\it classical} superposition of fields with opposite 
phases. At larger delays $T$, the plateau $P_{ee}=1/2$ turns to a 
slow decay to zero because the two coherent states of the mixture both 
tend to the vacuum state and start to overlap, due to field energy 
dissipation \cite{dav}.

This conditional probability decay can be experimentally 
reconstructed by sending a large number of atom pairs for each delay 
time $T$, obtaining therefore a clear observation of the 
decoherence phenomenon in its time development. Actually, 
in Ref.~\cite{prlha}, the experimental demonstration of decoherence 
has been given by considering not simply $P_{ee}$ but the difference 
between conditional probabilities $\eta = P_{ee}-P_{ge}$.

\section{The stroboscopic feedback model}

We now propose a modification of the experiment of 
Brune {\it et al}. \cite{prlha} in which the cat decoherence is not simply 
monitored but also controlled in an active way. The idea is to apply 
the same feedback scheme described above for optical cavities, which 
gives a photon back to the cavity whenever the photodetector clicks. 
However in this microwave case one has to find a different way to determine 
if the cavity mode has lost a photon or not, because there are no good 
photodetector available in this wavelength region. 
Ref.~\cite{prlha} suggests using
off-resonant atoms crossing the cavity to measure the cavity field and 
therefore
in this case one could replace continuous photodetection with a 
stroboscopic measurement performed by a sequence of off-resonant 
probe atoms, separated by a time interval $T$. A sort of indirect 
microwave photodetection can be obtained by using the fact that, as 
suggested by Eq.~(\ref{prob}), the detection of the $e$ or $g$ atomic 
level is equivalent to the measurement of the parity of the cavity 
mode state. In fact, Eq.~(\ref{offre}) for the conditioned
cavity mode density matrices $\rho_{g/e}$ can be rewritten in the 
following way
\begin{eqnarray}
\rho_{e} &=& P_{odd}\rho P_{odd}
\label{offre21} \\
\rho_{g} &=& P_{even}\rho P_{even} \;,
\label{offre22} 
\end{eqnarray}
where $P_{odd}$ ($P_{even}$) is the projector onto the subspace 
with an odd (even) number of photons and therefore finding the atom 
in the state $e$ ($g$) means measuring a parity $P=-1$ ($P=+1$) for 
the state of the microwave mode within the cavity $C$.

To fix the ideas, let us consider from now on the case when the cat 
state generated by the first off-resonant atom is an odd coherent 
state (first atom detected in $e$).  When a second probe atom crosses 
the cavities arrangement after a time interval $T$ and is detected in 
$e$, it means that the cavity mode state has remained in the odd 
subspace, or, equivalently, that the cavity has lost an {\it even} 
number of photons.  If the time interval $T$ is much smaller than the 
cavity decay time $\gamma^{-1}$, $\gamma T \ll 1$, then the 
probability of loosing two or more photons is negligible and one can 
say that finding the state $e$ means that no photon has leaked out 
from the high-Q cavity $C$.  On the contrary, when the probe atom is 
detected in $g$, the cavity mode state is projected into the even 
subspace and this is equivalent to say that the cavity has lost an 
{\it odd} number of photons.  Again, in the limit of enough closely 
spaced sequence of probe atoms, $\gamma T \ll 1$, the probability of 
loosing three or more photons is negligible and therefore finding the 
level $g$ means that one photon has exited the cavity.

\begin{figure}
\centerline{\psfig{figure=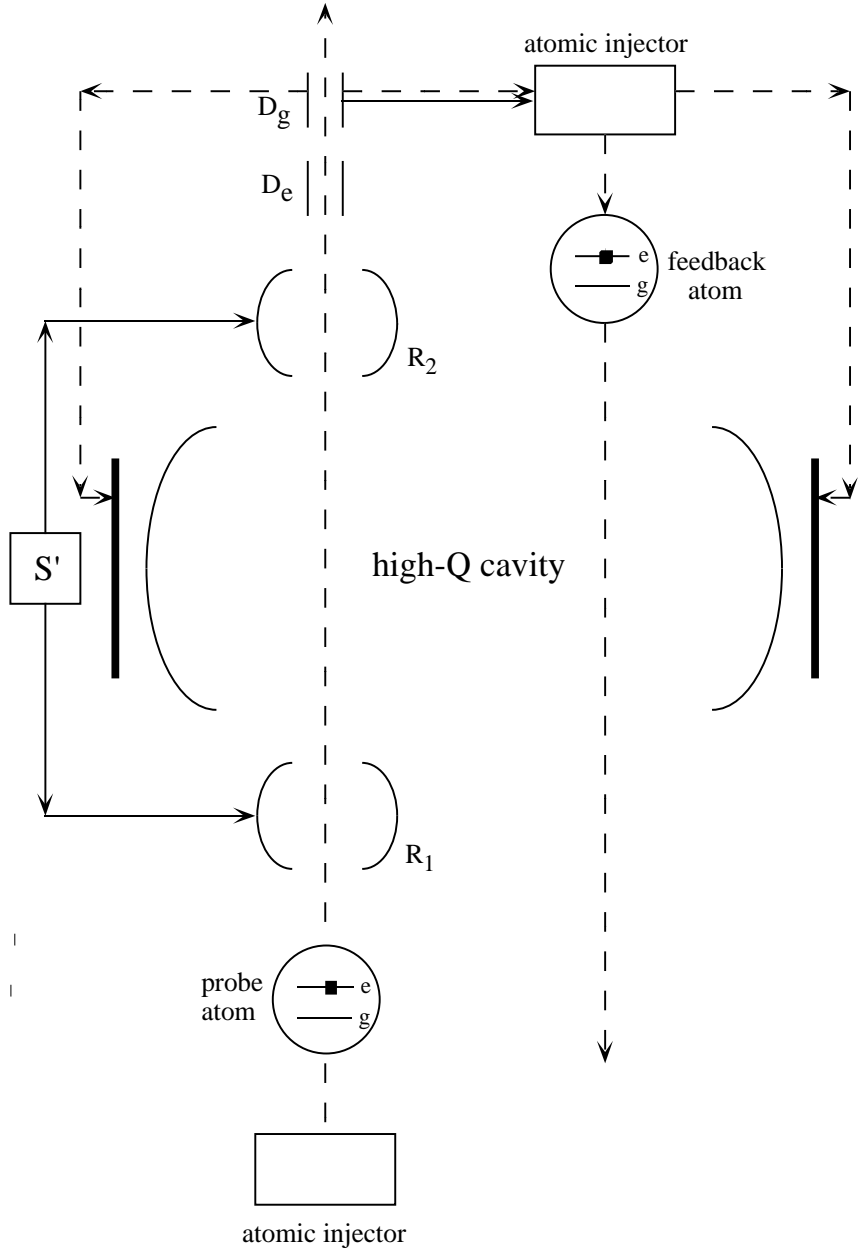,height=22cm}}
\caption{Schematic diagram of the stroboscopic feedback scheme
for the experiment of Brune {\it et al}. $R_{1}$ and $R_{2}$ are the two 
cavities in which classical microwave pulses can be applied. The feedback 
loop acts whenever the $g$-state detector clicks and it switches on both 
the atomic injector and the electric field in the high-Q cavity to 
Stark-shift the level $e$ into resonance.}
\label{appaha}
\end{figure}

Therefore, for achieving a good protection of the initial odd cat 
state, the feedback loop
has to supply the superconducting cavity with a photon whenever the 
probe atom is detected in $g$, while feedback must not act 
when the atom is detected in the $e$ state.
This feedback loop can be realized with a switch connecting the 
$g$ state field-ionization detector with another atom injector, sending 
an atom in the excited state $e$ into the high-Q cavity. 
This feedback atom has to be {\it resonant} with the radiation mode in the 
superconducting cavity and this can be obtained with another 
switch turning on an electric field in the cavity $C$ when the atom enters it, 
so that the level $e$ is Stark-shifted into resonance with the cavity mode.
A schematic representation of the experimental apparatus of 
Ref.~\cite{prlha} together with the feedback loop is given by 
Fig.~\ref{appaha}.

The time evolution of the microwave field in the high-Q cavity can be 
described stroboscopically by the transformation from the state just 
before the crossing of $n$-th non-resonant probe atom $\rho(nT)$,
to the state of the radiation 
mode before the next non-resonant atom crossing $\rho(nT+T)$. 
This transformation 
is given by the composition of two successive mappings:
\begin{equation}
\rho(nT+T) = 
\Phi(\rho(nT))=\Phi_{diss}\left(\Phi_{fb}(\rho(nT))\right)
\;, \label{mapp}
\end{equation}
where $\Phi_{fb}$ describes the effect of the interaction with the 
non-resonant atom followed by the effect of the resonant feedback 
atom, which interacts with the cavity field or not according to the 
result of the measurement performed on the off-resonant atoms. The 
operation $\Phi_{diss}$ describes instead the dissipative evolution 
of the field mode during the time interval $T$ between two successive 
atom injections and it is characterized by the energy relaxation rate 
$\gamma$. 

The feedback mechanism acts only on the density matrix $\rho_{g}$, 
conditioned to the detection of level $g$ and is described by the 
resonant interaction part of the Hamiltonian (\ref{jc})
\begin{equation}
H_{r}=\hbar \Omega \left(|e\rangle \langle g| a+|g\rangle \langle e| 
a^{\dagger}\right) \;;
\end{equation}
the effect on the cavity mode density matrix $\rho$ is then given by
(the feedback atoms are not detected after exiting the microwave 
cavity $C$)
\begin{equation}
\rho ' = {\rm 
Tr_{at}}\left\{\exp\left\{-\frac{i}{\hbar}H_{r}\tau \right\}\left(|e\rangle
\langle e|\otimes 
\rho 
\right)\exp\left\{\frac{i}{\hbar}H_{r}\tau \right\} \right\} \;,
\end{equation}
where $\tau$ is the interaction time of the feedback atom.
Performing the trace, one gets
\begin{eqnarray}
\label{firo}
&&\rho ' = 
\cos(\mu \sqrt{aa^{\dagger}})\rho \cos(\mu \sqrt{aa^{\dagger}}) \\
&&+a^{\dagger} (aa^{\dagger})^{-1/2}\sin(\mu \sqrt{aa^{\dagger}})\rho 
\sin(\mu \sqrt{aa^{\dagger}}) (aa^{\dagger})^{-1/2} a  \nonumber \;,
\end{eqnarray}
where $\mu =\Omega \tau $.
Then, we have to take into account the effect of the non-unit efficiency 
of the atomic detectors $\eta $, which is of the order of $\eta =0.4$ in 
the actual experiment. This means that the off-resonant atoms are not 
detected with probability $1-\eta $ and when this happens, the feedback 
loop does not act. Using both Eqs.~(\ref{offre}) and (\ref{firo}), 
we derive the 
explicit expression of the feedback operator $\Phi_{fb}$:
\begin{eqnarray}
\label{fb}
&&\Phi_{fb}(\rho) = \eta\rho_e +\eta\cos(\mu \sqrt{aa^{\dagger}})\rho_g 
\cos(\mu \sqrt{aa^{\dagger}})  \\
&&+\eta a^{\dagger} \frac{\sin(\mu \sqrt{aa^{\dagger}})}{(aa^{\dagger})^{1/2}}
\rho_g
\frac{\sin(\mu \sqrt{aa^{\dagger}})}{(aa^{\dagger})^{1/2}} a 
+(1-\eta)\left[\rho_e+\rho_g\right ] \;. \nonumber 
\end{eqnarray}
In writing this expression we have implicitely assumed that not only 
the off-resonant atom time of flight, but also the feedback loop delay 
time are much smaller than the typical timescales of the system and that 
they can be neglected. This assumption is essentially equivalent to 
the Markovian assumption made for the continuous photodetection 
feedback described above and it simplifies considerably the discussion. 

The operator $\Phi_{diss}$ describing the dissipative time evolution 
between two successive atom crossings can be obtained from the exact 
evolution of a cavity in a standard vacuum bath \cite{herzog} and it 
can be written as
\begin{equation}
\Phi_{diss}(\rho)=\sum_{k=0}^{\infty}A_{k}\rho A_{k}^{\dagger}\;,
\label{disso}
\end{equation}
where 
\begin{equation}
A_{k}=\sum_{n=0}^{\infty}\sqrt{\frac{(n+k)!}{n! k!} e^{-n \gamma 
T}\left(1-e^{-\gamma T}\right)^{k}} |n\rangle \langle n+k | \;. 
\label{cnk}
\end{equation}

If we now use the explicit expressions (\ref{fb}) and (\ref{disso}), 
we get the general expression of the transformation $\Phi$ of 
Eq.~(\ref{mapp}),
which can be written for the density 
matrix elements in the following way ($\langle n 
|\Phi(\rho)|n+p \rangle = \rho'_{n,n+p}$):
\begin{eqnarray}
&&\rho'_{n,n+p}=\sum_{k=0}^{\infty}\left\{\frac{c_{n,k}c_{n+p,k}}{4}
\left[\eta s_-(n,k)^{2} +4(1-\eta )
 \right. \right. \label{mapoele} \\
&&+ \eta s_+(n,k)^{2} 
\cos(\mu\sqrt{n+k+1})\nonumber \\
&&\left . \mbox{}\cos(\mu\sqrt{n+p+k+1})\right]
 \nonumber \\
&&+\eta \frac{c_{n,k+1}c_{n+p,k+1}}{4} s_{+}(n,k)^{2} 
\sin(\mu\sqrt{n+k+1})\nonumber \\
&&\left .\mbox{}\sin(\mu\sqrt{n+p+k+1}) \right\}
\rho_{n+k,n+p+k} \nonumber \\
&&+ \eta \frac{c_{n,0}c_{n+p,0}}{4}\sin(\mu \sqrt{n})\sin(\mu \sqrt{n+p})
s_{-}(n,0)^{2} \rho_{n-1,n+p-1} \nonumber \;,
\end{eqnarray}
where
\begin{eqnarray*}
c_{n,k} & = & \sqrt{\frac{(n+k)!}{n! k!} e^{-n \gamma 
T}\left(1-e^{-\gamma T}\right)^{k}}\\
s_\pm(n,k) & = &1\pm(-1)^{n+k} \;. 
\end{eqnarray*}

An important aspect of the above equation is that the time evolution of a 
given density matrix element depends only upon 
the matrix elements with the same ``off-diagonal'' index $p$. 
This implies in particular that only even values of $p$ can be 
considered in (\ref{mapoele}), because one starts from an odd coherent 
state and the matrix elements with $p$ odd, being zero initially, 
remain zero at any subsequent time. To state it in other words, 
if the initial state has a definite parity, the 
dynamical evolution is such that the cavity mode state evolves 
within the two subspaces with given parity and the projection into the 
space with no definite parity always remains zero. We have already 
used this fact in Eq.~(\ref{fb}) where we have written $\rho = 
\rho_{e}+\rho_{g}$, since, as showed by (\ref{offre21}) and 
(\ref{offre22}), these two matrices are just the odd and even 
components of the density matrix.

Generally speaking, the parity of the cavity mode state plays such a 
fundamental role that our stroboscopic feedback scheme is able to 
protect only even and odd coherent states (we have considered an initial 
odd cat state only, but the scheme can be simply adapted to the even case). 
In fact one could generalize the scheme described above and consider
the generation of more general cat states. For example, 
one can consider generic phase shifts $\phi\neq \pi $ 
(as it is done in \cite{prlha}) and generic microwave pulses in the two 
cavities $R_{1}$ and $R_{2}$
\begin{eqnarray}
|e\rangle & \rightarrow & c_{e}|e\rangle 
+c_{g} |g\rangle \nonumber \\
|g\rangle & \rightarrow & -c_{g}^{*}|e\rangle 
+c_{e}^{*} |g\rangle  \;,
\label{nopi2}
\end{eqnarray}
where $c_{e}$ and $c_{g}$ depend on the intensity and phase of the microwave
pulses in
$R_{1}$ and $R_{2}$ and on the interaction time. 
This allows to generate a large class of linear
superpositions of  coherent states with different phases, but only in the case
of cat 
states with a given parity our stroboscopic scheme can be implemented.
In fact the essential condition for the stroboscopic protection 
scheme to be applied is the existence of relations like (\ref{offre21}) and 
(\ref{offre22}) in which the cavity mode states conditioned to the 
detection of the two atomic levels are expressed as projections into 
given, orthogonal subspaces. Only in this case in fact, is it 
possible to correlate with no ambiguity one atomic detection with a 
state or property of the cavity mode and then consequently apply a 
feedback scheme. It is then easy to prove that the two microwave 
pulses in $R_{1}$ and $R_{2}$ and the dispersive interaction in $C$
(see Eq.~(\ref{2at2})) 
determine two projection operators only for the situation considered here, 
($\phi =\pi $ and two $\pi/2$ pulses) and these projectors are just 
the projectors into the even and odd subspace.

\section{Dynamics in the presence of stroboscopic feedback}

\begin{figure}
\centerline{\psfig{figure=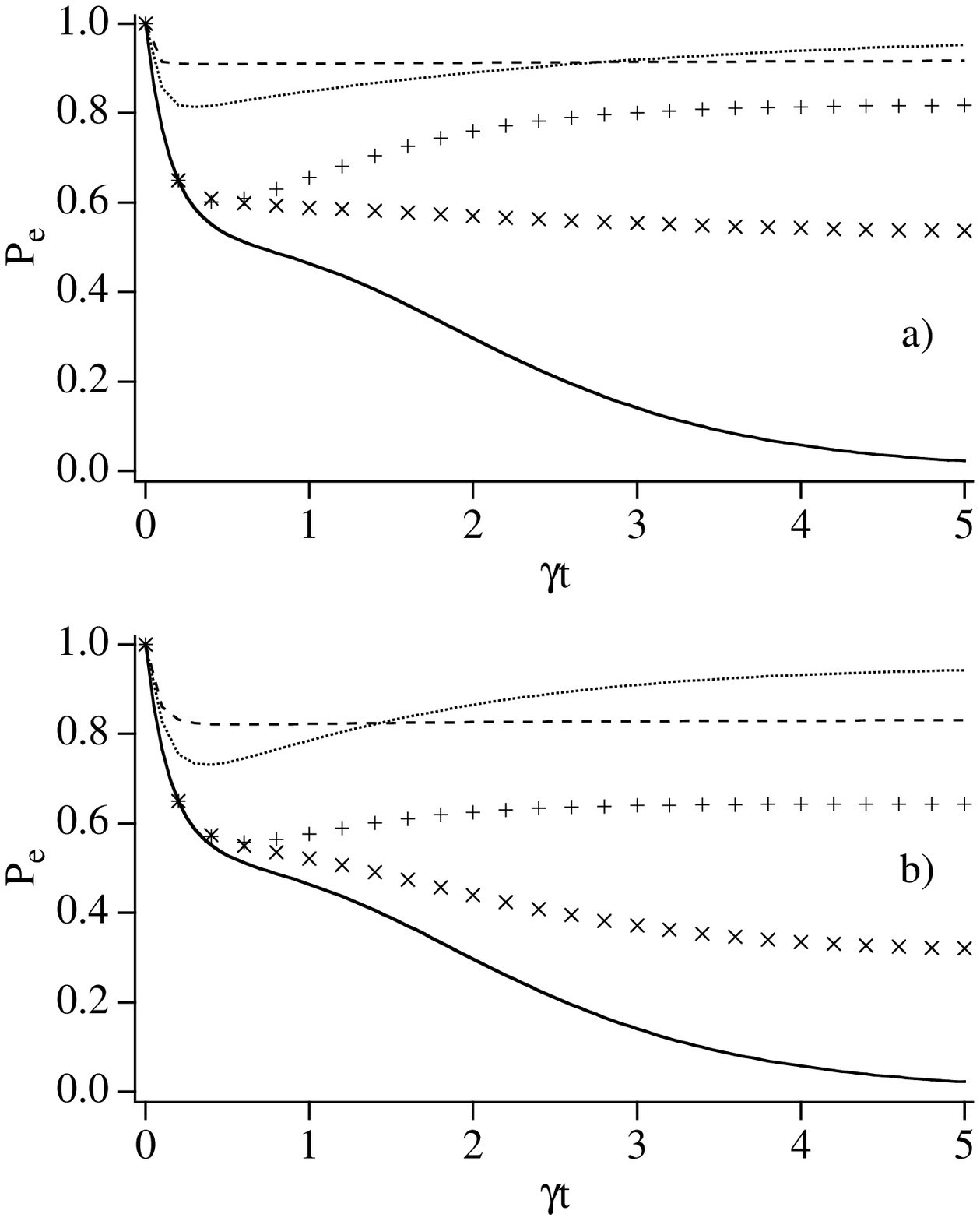,height=21cm}}
\caption{Time evolution of the probability of detecting the off-resonant 
atoms in the state $e$ in the case when $|\alpha |^{2}=3.3$. 
Full line: $\mu =0$ (no feedback case); 
dashed line: $\mu =\pi /6$ 
and $\gamma T=0.02$; dotted line: $\mu =\pi /2$ 
and $\gamma T=0.02$; horizontal crosses: $\mu =\pi /2$ 
and $\gamma T=0.2$; diagonal crosses: $\mu =\pi /6$ 
and $\gamma T=0.2$. In a) the ideal case of perfect atomic detection 
is considered, while b) refers to the case $\eta =0.4$}
\label{peef}
\end{figure}

The experimental study of this stroboscopic feedback scheme can be done 
performing a series of atomic detections of the state of the 
off-resonant probe atoms separated by a given time interval $T$ and 
repeating this series of measurements many times, always starting from 
a first detection in the state $e$. This allows one to reconstruct the 
time evolution of the probability of finding the state $e$,
$P_{e}(nT)$ (see Eq.~(\ref{prob})) in the presence of feedback.
The time evolution of this probability is plotted in Fig.~\ref{peef} where an 
initial odd coherent state with $|\alpha |^{2}=3.3$ (just the value 
corresponding to that of the actual experiment) is considered. The full 
line refers to the no feedback case ($\mu =0$), that is, the theoretical 
prediction of Eq.~(\ref{pee}); the dashed line refers to $\mu =\pi /6$ 
and $\gamma T=0.02$; the dotted line to $\mu =\pi /2$ 
and $\gamma T=0.02$; horizontal crosses to $\mu =\pi /2$ 
and $\gamma T=0.2$ and diagonal crosses to $\mu =\pi /6$ 
and $\gamma T=0.2$. In a) the ideal case of perfect atomic detection 
is considered, while b) refers to the case $\eta =0.4$, which is the 
actual efficiency of the detector employed in \cite{prlha}.
These two figures show the dependence on the three feedback 
parameters $\gamma T$, $\mu$ and $\eta$ and,  as  
expected, the most relevant one is the time between two 
successive measurements $T$. This time has to be as small as possible,
because decoherence 
can be best inhibited if one can ``check'' the cavity state, 
and try to restore it, as soon as possible. Moreover we 
have seen that the indirect measurement of the cavity with the atoms
becomes optimal only in the continuous limit $\gamma T \ll 1$ and only 
in this limit (and for ideal detection efficiency $\eta=1$) 
the initial photon number distribution is perfectly preserved.  

The coupling parameter $\mu = \Omega \tau$ is instead  
connected to the probability of releasing the 
photon within the high-Q cavity. We have assumed that the feedback 
atoms come from an independent source just to have the possibility
of varying their velocity and therefore the parameter $ \mu$.
This probability of releasing the photon in the cavity
is maximized when the sine term in 
(\ref{fb}) is maximum, i.e., when
\begin{equation}
\mu \sqrt{n}=\pi (m+1/2)\;\;\;\; m \;\;{\rm integer}
\label{sqreso}
\end{equation}
This resonance condition depends on the photon number $n$
which however is not determined in general and moreover decreases as time
evolves 
(when $\gamma T \neq 0$). In the case of the Schr\"odinger cat state 
studied here, (\ref{sqreso}) roughly corresponds to the condition 
$\mu |\alpha |=\pi (m+1/2)$ and this explains why at small times the 
case $\mu =\pi/6$ gives a good result ($|\alpha |^{2}=3.3$ in the figures). 
At longer times the value $\mu 
=\pi/2$ gives the better result and this is due to the fact that the 
cavity mean photon number has become approximately one. A complete 
explanation of the asymptotic behavior of the curves of Fig.~\ref{peef}
is given by the fact that, as long as $T \neq 0$, the 
stationary state of the cavity field is a mixture of the vacuum and the 
one-photon state, given by
\begin{eqnarray}
&&\rho^{stat}=\frac{e^{\gamma T}-1}{e^{\gamma 
T}-1+\eta \sin^{2}\mu }|0\rangle \langle 0| \nonumber \\
&& +\frac{\eta \sin^{2}\mu}{e^{\gamma 
T}-1+\eta \sin^{2}\mu}|1\rangle \langle 1| \;.
\label{statio}
\end{eqnarray}
It is immediate to see that this means
\begin{equation}
P_{e}(\infty)=\rho^{stat}_{11}=\frac{\eta \sin^{2}\mu}{e^{\gamma 
T}-1+\eta \sin^{2}\mu} \;.
\label{statio2}
\end{equation}
which is verified by the plots shown in Fig.~\ref{peef}.

The form of the stationary state can be obtained from the general 
expression of the mapping (\ref{mapoele}). In fact, since the time 
evolution of a given matrix element is coupled only to those with the 
same off-diagonal index $p$, this mapping can be written in the simpler 
form
\begin{equation}
\vec{V}'_{p} = A_{p} \vec{V}_{p} \;,
\label{sint}
\end{equation}
where $\rho _{n,n+p}$ is the $n$-th component of the vector 
$\vec{V}_{p}$ 
and $A_{p}$ is a matrix whose expression can be obtained from 
(\ref{mapoele}). The state of the cavity field after $K$ measurements 
(and eventual feedback corrections) is therefore obtained applying the 
matrix $A_{p}$ $K$ times. Since the evolution of the cavity field is 
dissipative, one can easily check that all the eigenvalues $\lambda$
of the family $A_{p}$ are such that $|\lambda | \leq 1$. The 
stationary state will correspond to the eigenvectors associated to the 
eigenvalue $\lambda =1$. It is possible to see that there is only one 
eigenvalue $\lambda =1$, for the matrix determining the evolution of the 
diagonal elements $A_{0}$, and that the associated eigenvector is the one 
corresponding to the diagonal stationary state of Eq.~(\ref{statio}). 

\begin{figure}
\centerline{\psfig{figure=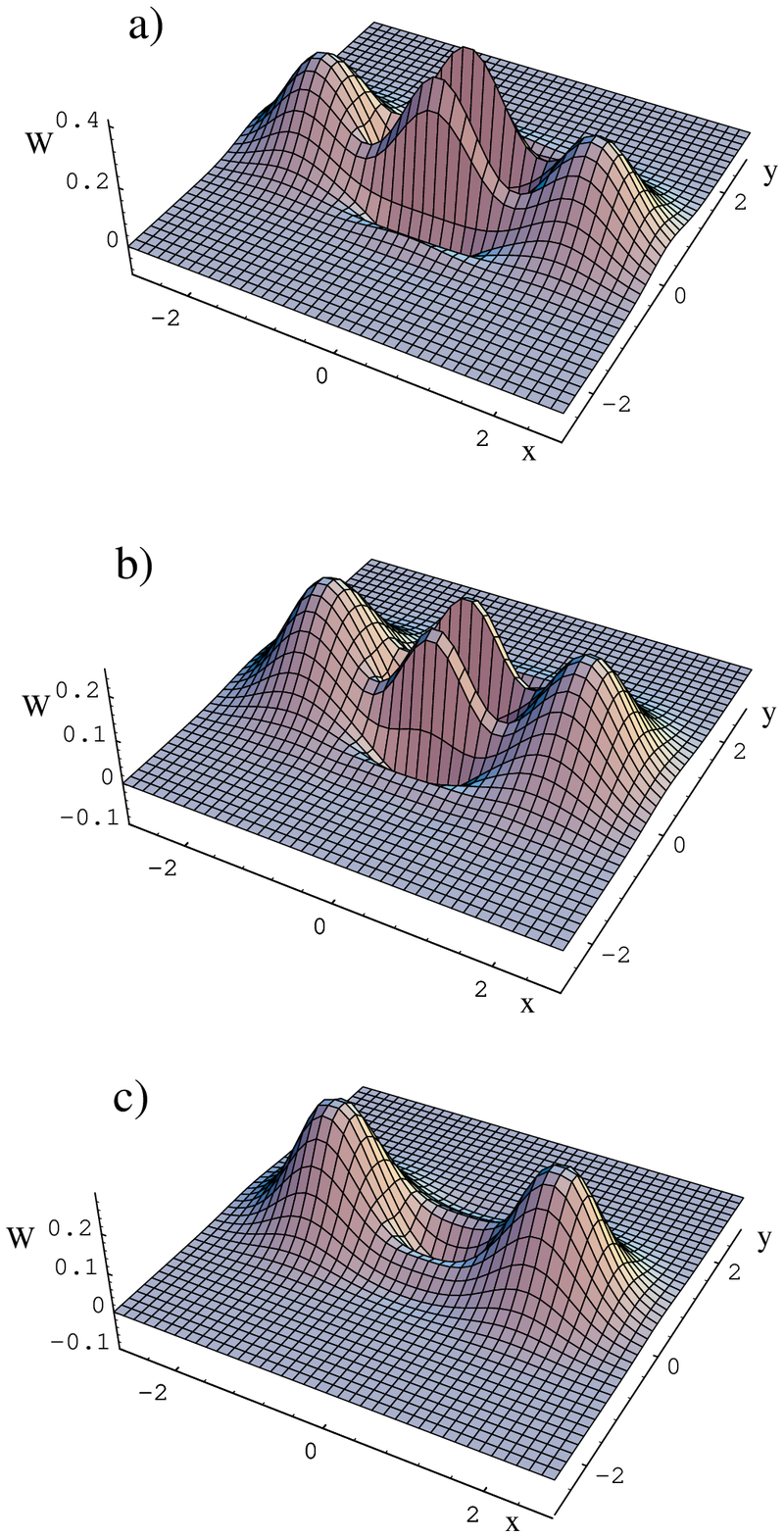,height=22cm}}
\caption{(a) Wigner function of the initial odd cat 
state, $|\psi \rangle = N_{-}(|\alpha \rangle - |-\alpha \rangle )$, $|\alpha 
|^{2}=3.3$ (b) Wigner function of the same cat state evolved for a 
time $t = 0.44/\gamma $ ($t\sim 3 t_{dec}$),
in the presence of feedback ($\mu =\pi /6$, $\gamma T=0.02$, $\eta=0.4$); (c) 
Wigner function of the same state after a time $t=0.44/\gamma$, 
but evolved in absence of feedback.}
\label{strobo1}
\end{figure}

At first sight, the comparison between the curves in the presence of 
feedback, with $P_{e}$ remaining close to one, and that in 
absence of feedback, seems to suggest that the initial odd cat state 
can be preserved almost perfectly. However, this is an incorrect 
interpretation because the quantity $P_{e}$ gives only 
a partial information on the state of the radiation mode within the 
cavity: it is a measurement of its parity (see Eq.~(\ref{prob})) 
and Fig.~\ref{peef} only shows that our feedback scheme 
is able to preserve almost perfectly the initial parity. 
Perfect cat state ``freezing'' can be 
realized only in cavities with an infinite $Q$; the proposed feedback 
scheme inevitably modifies the initial state, even in the ideal conditions of 
perfect detection efficiency $\eta =1$ and continuous feedback $\gamma T 
\approx 0$. In fact the stroboscopic feedback model shows the same 
behavior of the continuous feedback model discussed above for optical 
cavities, which (when restricting to initial states with given parity)
represents its continuous measurement limit $\gamma T \rightarrow 0$.
It is characterized by 
phase diffusion, because the photon left in the cavity by the resonant atom
has no phase relationship with those in the cavity. However this 
phase diffusion proves to be slower than the usual phase 
diffusion, so that also in this stroboscopic case, the protection of
the initial cat state is extremely good.
This is clearly shown by Fig.~\ref{strobo1},
where the Wigner function of the same initial odd coherent state 
considered in Fig.~\ref{peef},
is plotted in a) and compared with the Wigner function
of the cavity state after a time 
$t=0.44/\gamma$ ($t \sim 3 t_{dec}$) in the presence of feedback (b). 
The two states are almost indistinguishable, even if in 
Fig.~\ref{strobo1}b the actual experimental value $\eta =0.4$
is considered (the other parameters are $\mu =\pi /6$, $\gamma T=0.02$).
The comparison with Fig.~\ref{strobo1}c, where the Wigner function evolved 
for the same time interval in {\it absence} of feedback is plotted,
clearly shows 
the effectiveness of our scheme. Since $t\sim 3 t_{dec}$, the state in absence 
of feedback has become a mixture of two coherent states 
with opposite phases, and the oscillations associated to quantum 
coherence have essentially disappeared. 
On the contrary, the state evolved in presence of feedback is almost 
indistinguishable from the initial one and the interference 
oscillations are still very visible. Fig.~\ref{strobo1}b also shows that
the unconventional, feedback-induced phase diffusion is actually very 
slow, since its effects are not yet visible after $t\sim 3 t_{dec}$.

\begin{figure}
\centerline{\psfig{figure=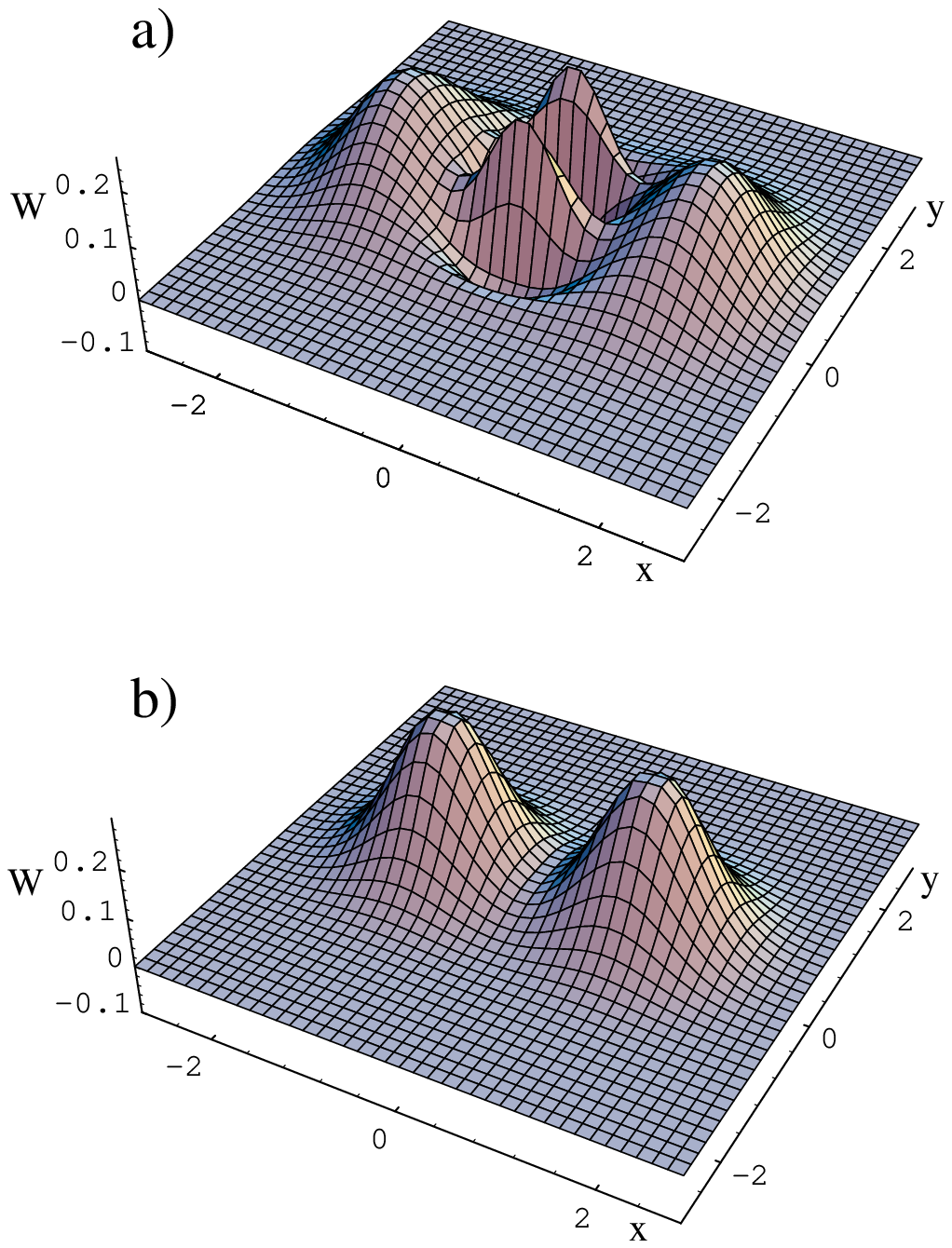,height=22cm}}
\caption{Wigner function of the odd cat state of Fig.~8,
evolved for a time $t = 1/\gamma $ in the presence of 
feedback with the same parameters as Fig.~8b (a),
and in absence of feedback (b).}
\label{strobo2}
\end{figure}

The effects of phase diffusion begin to be visible after one relaxation 
time $t=\gamma ^{-1}$, as shown by Fig.~\ref{strobo2}, where the Wigner 
functions at this time, both in the presence (a) and in absence (b) of 
feedback are compared (other parameter values are the same as in 
Fig.~\ref{strobo1}). Quantum coherence is quite visible in (a), while it has 
completely disappeared in (b); however the state in the presence of 
feedback begins to distort with respect to the initial state, as
the two peaks associated with the two coherent state become broader 
and more rounded due to phase diffusion.

\section{Concluding remarks}

In this paper we have presented a way for protecting a generic initial 
quantum state of a radiation mode in a cavity against decoherence. The 
initial quantum state is not perfectly preserved for an infinite time 
(this is possible only in a cavity with an infinite Q); nonetheless its 
quantum coherence properties can be preserved for a long time and the 
``lifetime'' of the state significantly increased.
The model presented here is a ``physical'' way to control decoherence 
based on feedback, that is, measuring the system and modify its dynamics 
according to the result of the measurement. In this sense it is very 
similar in spirit, to the proposals of Refs.~\cite{pell}.
Our approach is complementary to those based on quantum 
error correction codes \cite{error}, using software to deal with 
decoherence. 
The present feedback acts in 
a very simple way: one checks if the cavity has lost a photon, and when 
this happens, one gives the photon back through the injection of an 
appropriately prepared atom. In the case of a continuous monitoring of 
the system and in the ideal limit of unit detector efficiency, the model 
preserves perfectly the photon number distribution of the initial quantum 
state of the cavity. This is obtained at the price of introducing an 
unconventional phase diffusion, slower than the usual phase diffusion, (see 
Eqs.~(\ref{sqroot}) and (\ref{derise2})), that modifies the state at 
sufficiently long times. To be more specific, feedback 
protects very well the relative phase of the 
coefficients of the components of the initial state,
generating at the same time the diffusion of 
the phase of the field.

The above description of the feedback scheme explicitely considers all 
the experimental limitations (non-unit efficiency of the detectors, 
comparison between the various timescales) except one:
here we have assumed that one has an extemely good control of the atomic 
injection and that it is possible to send {\it exactly} one atom at a 
time in the cavity. This is not experimentally possible at the moment:
for example, in \cite{prlha} sending an ``atom'' explicitely means
sending an atomic pulse with an average 
number $\bar{n}\sim 0.2$, so that the probability of having 
two atoms simultaneously in the cavity is negligible. 
This fact makes the proposed feedback scheme much less effective; in fact, 
this is essentially equivalent to have, in the stroboscopic case, 
an effective quantum efficiency $\eta _{eff}=\eta  \bar{n}^{2}$, because
one has a probability $\bar{n}^{2} \sim 0.04$ of having one probe atom 
and one feedback atom in each feedback loop. As a consequence,
the dynamics in the presence of feedback becomes hardly 
distinguishable from the standard dissipative evolution.

In the continuous feedback scheme for optical cavities one has the  
feedback atomic beam only and the effective efficiency is $\eta \bar{n}$. 
Anyway in the optical case, the problem of having {\it exactly} one feedback 
atom at a time with certainty, could be overcome, at least in principle, 
replacing the beam of 
feedback atoms with a single fixed feedback atom, optically trapped by 
the cavity (for a similar configuration, see for example 
\cite{pell}). The trapped atom must have the same $\Lambda $ 
configuration described in Fig.~1 and the adiabatic photon transfer 
between the classical laser field and the quantized optical mode could be 
obtained with an appropriate shaping of the laser pulse $\Omega (t)$. The 
possibility of simulating the adiabatic transfer with an appropriately 
designed laser pulse has been recently discussed by 
Kimble and Law \cite{pistol} in a proposal for the realization of a 
``photon pistol'', 
able to release exactly one photon on demand (see also \cite{eberly}). 
In this case, the feedback loop would be simply activated by turning on the 
appropriately shaped laser pulse focused on the trapped atom. During the time 
interval between two photodetections, the atom 
has to remain in the ``ready'' state $|g_{1}\rangle $, 
which is decoupled from the cavity mode, and this could be obtained with 
an appropriate recycling process, driven, for example, by supplementary 
laser pulses \cite{pistol}.

\section{Acknowledgments}
This work has been partially supported by the Istituto Nazionale Fisica 
della Materia (INFM) through the ``Progetto di Ricerca Avanzata INFM-CAT''.

\newpage 

\newpage

\newpage

\newpage

\newpage

\newpage

\newpage

\newpage

\newpage

\end{document}